\documentclass[12pt]{article}
\pdfoutput=1

\usepackage[english]{babel}
\usepackage{epsf,amssymb,amsmath,bbold,bbm}
\usepackage{graphicx,color}

\numberwithin{equation}{section} 

\usepackage[colorlinks=true,       
    linkcolor=blue,          
    citecolor=blue,        
    filecolor=blue,      
    urlcolor=blue,   
    linktoc=page	
     ]{hyperref}

\usepackage{cite}

\makeatletter
\renewcommand\section{\@startsection {section}{1}{\z@}%
                               {-3.5ex \@plus -1ex \@minus -.2ex}
                               {2.3ex \@plus.2ex}%
                               {\normalfont\large\bfseries}}
\renewcommand\subsection{\@startsection{subsection}{2}{\z@}%
                                 {-3.25ex\@plus -1ex \@minus -.2ex}%
                                 {1.5ex \@plus .2ex}%
                                 {\normalfont\bfseries}}
\makeatother

\usepackage{xspace}

\usepackage{float}
\usepackage{caption}
\usepackage{subcaption}

\newcommand{\rd}{d}

\newcommand{\ads}{AdS$_3$\xspace}

\newcommand{\LF}{\left(}
\newcommand{\RF}{\right)}
\newcommand{\LT}{\left[}
\newcommand{\RT}{\right]}

\newcommand{\Tr}{\text{Tr}}
\newcommand{\pd}{\partial}
\newcommand{\Acal}{\mathcal{A}}
\newcommand{\vp}{\varphi}
\newcommand{\dv}{\dot{v}}
\newcommand{\dr}{\dot{r}}
\newcommand{\dvp}{\dot{\varphi}}
\newcommand{\la}{\lambda}
\newcommand{\Lcal}{\mathcal{L}}
\newcommand{\Ecal}{\mathcal{E}}
\newcommand{\Scal}{\mathcal{S}}
\newcommand{\Fcal}{\mathcal{F}}
\newcommand{\Tcal}{\mathcal{T}}

\newcommand{\atan}{\text{arctan}}
\newcommand{\atanh}{\text{arctanh}}
\newcommand{\jads}{J_{\text{AdS}}}
\newcommand{\eads}{E_{\text{AdS}}}

\newcommand{\phim}{\varphi}
\newcommand{\phip}{\bar{\varphi}}
\newcommand{\tm}{t}
\newcommand{\tp}{\bar{t}}
\newcommand{\rmin}{r}
\newcommand{\rp}{\bar{r}}

\newcommand{\bs}{\bar{s}}
\newcommand{\bt}{\bar{t}}

\newcommand{\e}{\epsilon}

\newcommand{\VBTZ}{$\mathcal{V}$BTZ }

\begin{document}

\begin{titlepage}

\begin{center}
{\Large \bf  Entanglement entropy in inhomogeneous quenches in \ads /CFT$_2$ }\\

\vskip 10mm

{\large Tim De Jonckheere$^{a}$, Jonathan Lindgren$^{b}$\\
\vspace{3mm}
}

\vskip 7mm

$^a$ Theoretische Natuurkunde, Vrije Universiteit Brussel, and \\ International Solvay Institutes,
Pleinlaan 2, B-1050 Brussels, Belgium \\
$^b$Nordita, KTH Royal Institute of Technology and Stockholm University,
Roslagstullsbacken 23, SE-106 91 Stockholm, Sweden \\

\vskip 6mm
{\small\noindent  {\tt Tim.De.Jonckheere@vub.be, jonathan.lindgren@su.se}}

\end{center}
\vfill

\begin{center}
{\bf ABSTRACT}\\
\end{center}
We compute entanglement entropy and differential entropy in inhomogeneous holographic quenches in \ads /CFT$_2$. The quenches are arbitrarily inhomogeneous and modeled by an infalling shell of massless non-rotating matter where the final state is not dual to a static black hole but rather to a black hole with time-dependent stress-energy tensor modes. We study the entanglement entropy of an interval and differential entropy of a family of intervals analytically when the inhomogeneities have a perturbative amplitude and numerically for non-perturbative inhomogeneities. While we are in principle able to study these quantities for any inhomogeneities, we discuss two concrete examples: an oscillatory quench and a bilocal quench. Both cases display saturation towards a steady state but do not fully thermalize. Depending on the location and size of the interval, the entanglement entropy displays a variety of interesting phenomena such as plateau phases, bumps, and discontinuities in its first derivative with respect to time. 
\vspace{3mm}
\vfill
\end{titlepage}

\tableofcontents

\section{Introduction}
When interacting systems are brought out of equilibrium they tend to equilibrate to a state that effectively looks thermal. Exactly how this happens is a central theme of current research and varies from system to system. In case of a far-from-equilibrium initial state it is not known in general how the system evolves to thermal equilibrium. One way to model the evolution involves quantum quenches, e.g.\,,~\cite{Calabrese:2006rx,Calabrese:2007rg}, where a sudden energy injection effectively brings the system to a highly excited state. By spontaneous thermalization the system relaxes to equilibrium. Quenches are typically implemented by a sudden change in the Hamiltonian. They have been especially interesting in $d=1+1$ conformal field theory (CFT) because of the high degree of analytic control over the model. Quantum quenches have interesting applications in physical systems  such as Heisenberg spin chains, Luttinger liquids, as in~\cite{Cazalilla:2006htn,Sotiriadis:2015xia} or one-dimensional ultracold atomic gases, e.g.\,,~\cite{Geiger:2014,Piroli:2016,Caux:2013ra,Kormos:2013xta}. Most results in conformal field theory have focused on homogeneous global quenches (where the quench is spatially translation invariant) and local quenches like in~\cite{Calabrese:2007mtj,Asplund:2014coa}. Far less is known about thermalization under inhomogeneous global quenches. Nevertheless, they are actually the more interesting case because spin chains and 1D gases are typically spatially varying due to a non-uniform density or a spatially varying potential. Inhomogeneous quenches have been studied in limiting cases of $d=1+1$ CFTs where the inhomogeneities are perturbatively small compared to an average value~\cite{CardySotiriadis2008}, in free bosonic field theories e.g.\,,~\cite{Dubail:2016tsc} and integrable models e.g.\,,~\cite{Alba:2017qag}. Two-dimensional conformal field theories have an infinite number of conserved charges and it is a fascinating question to what extent such theories thermalize. Left- and right-moving modes are decoupled and conserved, so one generically does not expect a full equilibration.\\
Ultimately, one would be interested in studying quenches in a regime of strong interactions. In this paper, we study $d=1+1$ inhomogeneous quenches directly at large coupling via AdS/CFT, a route which has been considered before in for example~\cite{Balasubramanian:2013oga} in higher dimensions and in e.g.\,,~\cite{Erdmenger:2017gdk,Arefeva:2017pho} in \ads /CFT$_2$.   These works treated cases where the scale of spatial variation is small or focused on particular quench profiles. We make use of a new type of holographic quench introduced in~\cite{Lindgren:2015fum,Lindgren:2016wtw}, for any type of quench density profile. The holographic spacetime is a Vaidya-like geometry with an inhomogeneous massless shell inserted. Homogeneous Vaidya has been studied in~\cite{Balasubramanian:2011ur,Hubeny:2013dea,Ziogas:2015aja,AbajoArrastia:2010yt} and is dual to a CFT state formed by a ring of local operators~\cite{Anous:2016kss}. The inhomogeneous Vaidya geometry can be viewed as a limit of a large number of collapsing massless point-particles in \ads. The final geometry after injection of the shell is that of a BTZ black hole on which large gauge transformations\footnote{Proper gauge transformations are gauge transformations that approach the identity of the asymptotic symmetry algebra, while large (improper) gauge transformations approach a non-trivial element of the asymptotic symmetry algebra and these map physical states into new physical states, see for instance \cite{Regge:1974zd,Benguria:1976in,Bunster:2014mua}.} are applied and which we call \VBTZ (for Virasoro BTZ). These are black hole solutions which have a non-trivial (time-dependent) boundary stress-energy tensor. \\
As a diagnostic of thermalization we study the dynamics of holographic entanglement entropy of an arbitrary interval $A$ of size $\ell$, as given by the Hubeny-Rangamani-Takayanagi (HRT) formula~\cite{Hubeny:2007xt}, which is a generalization to time-dependent situations of the well known Ryu-Takayanagi (RT) formula~\cite{Ryu:2006bv} proposed for static setups. In three dimensions, these prescriptions both state that one should compute the length of the geodesic with minimal length that connects to the two endpoints of the boundary interval $A$, and henceforth we will refer to this method as the HRT prescription. Homogeneous holographic quenches have shown a monotonic growth of entropy  and a saturation as soon as $t= \ell/2$. In this paper we show that inhomogeneous holographic quenches also display a smooth interpolation from AdS to \VBTZ\hspace{-2mm}, also during the time interval $0<t<\ell/2$ as we show in appendix \ref{app:saturation}. From $t=\ell/2$ onwards the entropy behaves as in eternal \VBTZ\hspace{-1.7mm}. Because this is not a static geometry however, the late time behaviour of the holographic entropy can be very non-trivial. \\
A second diagnostic is differential entropy~\cite{Balasubramanian:2013lsa,Headrick:2014eia}. This is the residual entropy of a continuous family of observers who can each perform measurements on an interval $A_\zeta$ where $\zeta$ parametrizes the family. This quantity is  more broadly interesting for the purpose of bulk reconstruction in AdS/CFT. One might wonder to what extent the bulk geometry can be reconstructed from holographic entanglement entropy (as measured by the HRT formula). As we will review in section \ref{sec:review_differential_entropy}, the differential entropy reproduces the length of a closed bulk curve. If all closed bulk curves can be reproduced from differential entropy, then one might loosely say that the bulk is reconstructable from entanglement in the boundary CFT. It has been shown in~\cite{Balasubramanian:2013lsa} that some static geometries contain finite bulk regions that cannot be probed by differential entropy. These regions are called entanglement shadows. While anti-de Sitter space is reconstructable from entanglement in this respect, BTZ black holes have finite entanglement shadows outside the horizon. In situations where black holes are formed from collapse, one might ask whether entanglement shadows are produced dynamically. In this paper we discuss how they are produced both in homogeneous Vaidya and in the inhomogeneous shell background.\\

In section~\ref{sec:shell} we review the geometry of a massless inhomogeneous collapsing shell and the resulting \VBTZ geometry. In section~\ref{sec:EE} we review holographic entanglement entropy and differential entropy. Holographic entanglement entropy in two dimensional holographic conformal field theory follows by studying the length of geodesics in the gravity dual. We review the construction of geodesics with equal time endpoints on the boundary in the following well known geometries: global static BTZ,  global \ads $ $ and global Vaidya. Because these are well known cases, we will only collect and present formulae that are relevant for studying the entropy in \VBTZ $ $ and in the inhomogeneous shell background. We will first construct geodesic solutions in static BTZ and in \ads. Then we will review the computation of entropy in these two geometries for an equal time interval on the boundary. After that, we will compute geodesics and entropy in the Vaidya geometry. Sections~\ref{sec:perturbative} and~\ref{sec:non_perturbative} are devoted to the derivation of entanglement entropy and differential entropy in the \VBTZ geometry and the inhomogeneous shell. We do this for perturbatively small inhomogeneities in section~\ref{sec:perturbative} and non-perturbative ones in section~\ref{sec:non_perturbative}. Finally, in section~\ref{sec:examples} we study the evolution of entropy for two specific profiles of the shell. In one example, the density profile of the quench is oscillatory on top of an average density while in the other example we study the extreme limit of two local spikes in the energy density on the vacuum, in such a way that the total energy is above the black hole threshold. We conclude in section~\ref{sec:conclusions}.

\section{An inhomogeneous collapsing shell in AdS}\label{sec:shell}
In this section we describe the construction and properties of an inhomogeneous thin shell of lightlike matter collapsing to form a black hole solution. Such a spacetime is dual to an instantaneous, but not necessarily rotationally symmetric, quench in the boundary field theory, as we shortly motivate. As was shown in \cite{Lindgren:2015fum,Lindgren:2016wtw}, we can construct such a solution by taking the limit of an infinite number of massless particles that fall in from the boundary. We refer to \cite{Lindgren:2015fum,Lindgren:2016wtw} for details of this construction and here we focus instead on the final result. The resulting spacetime can then be formulated in a manner similar to the homogeneous AdS$_3$-Vaidya spacetime. We define the two metrics
\begin{equation}
d\bs ^2=-(\rp^2-M)d\bar v ^2+2d\bar v d\rp+\rp^2d\phip^2, \quad ds ^2=-(r ^2+1)dv ^2+2dv dr +r ^2d\phim^2,\label{metrics}
\end{equation}
where the first one is that of a BTZ black hole and the second one is that of \ads, both in infalling coordinates. For later reference, we will also state the metrics in global time coordinates, given by
\begin{equation}
\rd \bs^2=-(\rp^2-M) \rd \bt^2+\frac{\rd\rp^2}{\rp^2-M}+\rp^2\rd \phip^2,\label{BTZ_metric}
\end{equation}
for the BTZ black hole metric, obtained by the coordinate transformation $d\bar{v}=d\bt+d\rp / (-M+\rp^2)$, and for \ads we have
\begin{equation}
ds^2 = -\LF \rmin^2 +1\RF d\tm^2 + \frac{d\rmin^2}{\rmin^2 + 1} +\rmin^2 d\phim^2,\label{eq:ads_metric}
\end{equation}
obtained by the coordinate transformation $dv = dt + d\rmin /(\rmin^2+1)$. We have for convenience set the AdS radius to one, which will be the case throughout this paper. The parameter $M$ sets the mass of the final black hole solution and the event horizon (of the final black hole spacetime, not of the full spacetime) is given by $r_H=\sqrt{M}$. These two metrics are glued together along the surface $\bar v = v=0$ which defines the massless shell. The reader may note that this looks exactly like AdS$_3$-Vaidya if the coordinates $(\bar v,\rp,\phip)$ and $(v,r,\phim)$ are continuously related across the shell. However, in our case they are not necessarily continuous, instead when crossing the shell they are related by a specific coordinate transformation that reads
\begin{equation}
\phip=F(\phim),\quad \rp=r /F'(\phim).\label{coord_tr}
\end{equation}
The function $F$ completely specifies the spacetime and explicitly breaks the rotational symmetry and for this coordinate transformation to be well defined we demand that $F'>0$. Note that the coordinate $v$ is continuous across the shell and therefore we will drop the bar on $\bar v$ from now on. These solutions are the natural inhomogeneous generalizations of the thin shell AdS$_3$-Vaidya spacetime. The coordinate transformation \eqref{coord_tr} is also consistent in the sense that the induced metric on the shell at $v=0$, which is given by $ds_{\text{shell}}^2=r^2 d\phim^2$ and $d\bs_{\text{shell}}^2=\rp^2 d\phip^2$, is the same from both sides of the shell, namely $d\bs_{\text{shell}}^2=ds_{\text{shell}}^2$. This is a non-trivial consistency check and is necessary if we for example want to analyze this solution using the junction formalism in general relativity \cite{Israel:1966rt,Barrabes:1991ng}.\\

In the coordinates \eqref{metrics} it looks like the final spacetime is a static BTZ black hole. However, this is just an artifact of the coordinates we are using. In the present coordinate system, the spacetime is not manifestly asymptotically AdS, and doing a coordinate transformation to bring it to an asymptotically AdS spacetime would require performing a large gauge transformation which then would change the interpretation of the final state. The final state is instead a dressed BTZ black hole (\VBTZ), or in other words, a BTZ black hole on which we have applied a large gauge transformation where the stress-energy tensor modes are not constant. We expand more on this in Section \ref{stress_energy_CFT}.\\

\subsection{Stress-energy tensor of the dual CFT}\label{stress_energy_CFT}

In this section we compute the stress-energy tensor in the dual CFT. Part of the material here can also be found in \cite{Lindgren:2016wtw}. Note that the metrics before and after the collapsing shell are currently given by the metrics in \eqref{metrics}.  However, these coordinates are not continuous at the boundary and thus the spacetime is not asymptotically \ads. We now do a coordinate transformation such that the coordinates become continuous across the shell at the boundary and the spacetime is everywhere asymptotically \ads. This is a large gauge transformation and thus changes the asymptotic behaviour of the metric at the boundary and it is in these coordinates we should identify the boundary CFT stress-energy tensor modes, which will in general be non-constant. To be more specific, we will find a coordinate transformation from the metric \eqref{BTZ_metric} to
\begin{equation}
\rd \bs^2=\rd \rho^2+ T_+(y_+) (dy_+)^2+ T_-(y_-) (dy_-)^2+(e^{2\rho}+ T_+(y_+) T_-(y_-)e^{-2\rho})\rd y_+\rd y_-,\label{inhom_BTZ_metric}
\end{equation}
with the requirement that the coordinates above the shell are continuously connected to the coordinates below the shell at the boundary. The condition that the coordinates are continuous at the boundary will uniquely fix the functions $T_\pm$ in \eqref{inhom_BTZ_metric}. 

The metric \eqref{inhom_BTZ_metric} describes a class of asymptotically \ads solutions of three-dimensional gravity with negative cosmological constant with arbitrary boundary stress-energy tensor modes\cite{Banados:1998gg}. We only construct this coordinate transformation in an asymptotic expansion close to the boundary. Starting with the metric \eqref{BTZ_metric}, we first do the change of variables 
\begin{equation}
\phip=(x_++x_-)/2,\quad \bt=(x_+-x_-)/2, \quad  \rp=R+R^{-1} M/4,\label{eq:r_to_R}
\end{equation}
to obtain
\begin{equation}
\rd \bs^2=R^{-2}\rd R^2+ \frac{M}{4} ((dx_+)^2+ (dx_-)^2)+(R^2+ \frac{M^2}{16}R^{-2})\rd x_+\rd x_-.\label{metric2}
\end{equation}
This is the BTZ black hole metric written in the form \eqref{inhom_BTZ_metric} with $T_\pm=M/4$. Note that this is not a large gauge transformation and the spacetime is still interpreted as a static BTZ black hole. To generate spacetimes with general stress-energy tensor modes, we now do a change of coordinates $R=R(\rho,y_\pm)$, $x_\pm=x_\pm(\rho,y_\pm)$ such that the metric remains on the form \eqref{inhom_BTZ_metric} but now instead with some arbitrary functions $T_\pm(y_\pm)$. Such coordinate transformation is a large gauge transformation and the resulting spacetime is thus not anymore interpreted as a static BTZ black hole. We will restrict to the case where $T_\pm$ take the same functional form, namely $T_\pm(y_\pm)\equiv T(y_\pm)$ (this corresponds to a shell without angular momentum). When gluing this spacetime to the \ads spacetime across the shell at $v=0$, the function $T$ will be fixed by enforcing that the boundary coordinates are continuous when crossing the shell (recall that the original coordinates are discontinuous according to \eqref{coord_tr}). 
We only compute this coordinate transformation in an expansion in $e^\rho$. We thus write
\begin{align}
R=&R^{\LT 0\RT}(y_+,y_-) e^\rho+R^{\LT 1\RT}(y_+,y_-) e^{-\rho}+R^{\LT 2\RT}(y_+,y_-) e^{-3\rho}+\ldots,\\
x_\pm=&F(y_\pm)+x_\pm^{\LT 1\RT}(y_+,y_-) e^{-2\rho}+x_\pm^{\LT 2\RT}(y_+,y_-) e^{-4\rho}+\ldots,\label{eq:x_coord_tr}
\end{align}
where we define $x_\pm^{\LT 0\RT}\equiv F$ and all other coefficients are computed in terms of $F$. The fact that $x_\pm^{\LT 0\RT}$ only depends on one coordinate, as well as the functional form of $R^{\LT 0\RT}$ that we show below, can be immediately deduced by just assuming that the boundary metric remains flat. Furthermore, the choice $x_\pm^{\LT 0\RT}=F$ is imposed such as to ensure that the angular coordinate is continuous on the boundary when crossing the shell since in the original coordinate system they are related by \eqref{coord_tr}. A straightforward calculation gives the other functions in the expansion as
\begin{equation}
R^{\LT 0\RT}=(F'(y_-)F'(y_+))^{-1/2},\quad R^{\LT 1\RT}=\frac{F''(y_-)F''(y_+)}{\sqrt{F'(y_-)F'(y_+)}^3},\label{R_coord_tr}
\end{equation}
\begin{equation}
x_\pm^{\LT 1\RT}=-\frac{F''(y_\mp)F'(y_\pm)}{2F'(y_\mp)} ,\quad x_\pm^{\LT 2\RT}=\frac{(F''(y_\mp))^2F''(y_\pm)}{8(F'(y_\mp))^2}+\frac{M}{8}(F'(y_\mp))^2F''(y_\pm).
\end{equation}
 After carrying out this coordinate transformation, the function $T$ can be easily read off from the metric and equals
\begin{equation}
T=\frac{M}{4}(F')^2-\frac{F'''}{2F'}+\frac{3(F'')^2}{4(F')^2}.\label{stress_energy_CFT_eq}
\end{equation}
We can thus parametrize the solutions either in terms of $T$ or in terms of $F$. Notice that $T$ is in fact the CFT stress-energy tensor up to a factor of $4G_N$, so whenever we refer to the stress-energy tensor in this paper, we really just mean $T$.\\

In later sections we consider computing entanglement entropies of static intervals, so it is useful to consider how a static interval is affected by the coordinate transformations in this section. Note that the correct interpretation is that the endpoints of the interval are constant in the coordinates with metric \eqref{inhom_BTZ_metric} which are continuously connected to the coordinates below the shell at the boundary. To be more explicit the endpoints of a static interval $(\vp^{(1)},\vp^{(2)})$ at some time $t$ is given in the coordinates $y_\pm$ in \eqref{inhom_BTZ_metric} by $y_{+}^{(1)}+y_{-}^{(1)}=2\vp^{(1)}$ and $y_{+}^{(2)}+y_{-}^{(2)}=2\vp^{(2)}$, with $y_{+}^{(1)}-y_{-}^{(1)}=y_{+}^{(2)}-y_{-}^{(2)}=2t$. In the BTZ coordinates \eqref{BTZ_metric}, which we use for most of our calculations, the interval endpoints are then located at $x_{\pm}^{(i)}=F(y_{\pm}^{(i)})$ with $i=1,2$ as can be seen from \eqref{eq:x_coord_tr}. 

Moreover, when evaluating entanglement entropies in asymptotically AdS spacetimes it is necessary to introduce a cutoff close to the AdS boundary. The cutoff will also transform according to the coordinate transformation \eqref{R_coord_tr} and it is in the coordinates with metric \eqref{inhom_BTZ_metric} where we will assume that the cutoff is at a constant radial location $\rho_c$. Thus in the coordinate system \eqref{BTZ_metric}, where we do most of our computations, the cutoff will not be constant. Note however that the important point is not that the cutoff is constant in the coordinate system \eqref{inhom_BTZ_metric}, in principle the cutoff can be taken to be both time and space dependent also in this coordinate system. The important point is that we use the same cutoff when comparing the result from two different states, and that this cutoff is defined in the coordinates \eqref{inhom_BTZ_metric}. In that case any time and space dependencies of the cutoff will cancel out. To be more explicit, the equation \eqref{R_coord_tr} together with the relation \eqref{eq:r_to_R} between the radial coordinates $\bar{r}$ and $R$, can be used to relate the cutoff $\rho_c$ to the cutoffs $\bar{r}_{c}^{(i)}$ for the two endpoints $i=1,2$, and the result is
\begin{equation}
\rp_{c}^{(i)}=R_{c}^{(i)}+O(e^{-\rho_c})=\sqrt{F'(y_{-}^{(i)})F'(y_{+}^{(i)})}e^{\rho_c}+O(e^{-\rho_c}),\label{eq:r_cutoff}
\end{equation}
where the higher order terms are irrelevant for the results of the entanglement entropies computed in the $\rho_c\rightarrow\infty$ limit.\\

It is also possible to compute the spacetime stress-energy tensor, as it appears in the right-hand side of Einstein's equations, by using the junction formalism\cite{Israel:1966rt,Barrabes:1991ng} of general relativity. This stress-energy tensor is that of an infinitesimal shell of pressureless lightlike matter, as expected, with energy density profile proportional to that of the final \VBTZ spacetime (given by \eqref{stress_energy_CFT_eq}) minus the constant energy density in the initial \ads spacetime. As these results are not relevant for us in this paper we will not go through the procedure here, but instead refer to \cite{Lindgren:2015fum,Lindgren:2016wtw} for more details.

\section{Review of holographic entanglement entropies and geodesics}\label{sec:EE}
A very useful probe of correlations is the entanglement entropy. An observer typically only has access to a local subsystem $A$ and its entanglement entropy measures the amount of information that can be learned about the complementary subsystem $\bar{A}$ by performing measurements on $A$, at least when the system is in a pure state. In any case, the entropy is computed as the von Neumann entropy
\begin{equation}
S\LF \rho_A\RF = -\Tr\LF \rho_A\log\rho_A\RF,
\end{equation}
of the reduced density matrix $\rho_A$ on $A$. Entanglement entropy has been an intensively studied quantity in field theory over the past fifteen years especially in view of the holographic duality. In~\cite{Ryu:2006bv,Hubeny:2007xt}, it was conjectured that the von Neumann entropy of the reduced density matrix on a subsystem $A$ in a holographic field theory is computed by the area $\mathcal{A}$ of an extremal codimension two spatial surface that ends on the boundary $\pd A$ of $A$ and is homologous to $A$,
\begin{equation}
S\LF \rho_A\RF = \frac{\Acal\LF A\RF}{4G_N},\label{eq:RT}
\end{equation}
and we refer to this as the HRT prescription from the names of the authors of \cite{Hubeny:2007xt}. If many such surfaces exists, the one with smallest area should be chosen \cite{Ziogas:2015aja}. In gravity in $2+1$ dimensions the codimension two surfaces are spatial geodesics. \\

In this section we review holographic entanglement entropy in various backgrounds: a static BTZ black hole, \ads and \ads -Vaidya.  We will first construct the corresponding geodesics and then compute their lengths when anchored on an interval of size $\ell$ on the asymptotic boundary. Secondly, we compute differential entropy and comment on the existence of entanglement shadows in \ads and BTZ. We postpone the discussion of differential entropy in \ads -Vaidya to section \ref{sec:shell_perturb} because we wish to clearly separate review chapters from new results. 
\subsection{Spatial geodesics in the BTZ black hole}\label{sec:btzgeo}
In this section we will study spatial boundary anchored geodesics in the BTZ black hole background. Because the geodesic length of such a geodesic diverges near the asymptotic boundary, we will introduce a cutoff surface which we will assume here to lie at a fixed radius $r=r_c$. We will need geodesics in the BTZ black hole background later in this paper for two reasons. Firstly, this background serves as a zeroth order version of the \VBTZ $ $ geometry with perturbative inhomogeneities. Secondly, we can use the formulae for geodesics in the BTZ background also for the study of the non-perturbative \VBTZ $ $ geometry, because there exists a coordinate transformation to a metric in which the \VBTZ $ $ geometry looks like a static black hole. In the latter case the cutoff surface would not be at constant radius in those coordinates, but we will only deal with these issues later in this paper and just review the standard construction of geodesics in the static BTZ background here. The BTZ black hole background in infalling coordinates is given by the metric
\begin{equation}
ds^2 = -\LF r^2-r_H^2\RF dv^2 + 2dvdr + r^2d\vp^2,
\label{eq:metric_btz_review}
\end{equation}
where $r_H= \sqrt{M}$ with $M$ the black hole mass and where $\vp$ is an angular coordinate with periodicity $2\pi$. The $v$-coordinate is lightlike. Here we follow the conventions of~\cite{Balasubramanian:2011ur}. The geodesic equations of motion in BTZ are obtained by a straightforward variation of the spatial geodesic action
\begin{equation}
\mathcal{I}= \int\limits_{\la_1}^{\la_2} d\la \sqrt{ -\LF r^2-r_H^2\RF \dv^2 + 2\dv\dr + r^2\dvp^2},
\end{equation} 
where the dot stands for differentiation with respect to the affine parameter $\la$. Because of reparametrization invariance the affine parameter can always be chosen to be the eigenlength. Under this assumption the geodesic equations of motion are
\begin{eqnarray}
\dvp &=& \frac{r_H J}{r^2},\label{eq:EOM_phi}\\
\dr^2 &=& r_H^2 E^2 + \LF r^2-r_H^2\RF\LF 1-\frac{J^2 r_H^2}{r^2}\RF,\label{eq:EOM_r}\\
\dv &=& \frac{r_H E +\dr}{r^2-r_H^2},\label{eq:EOM_v}
\end{eqnarray}
where $E$ and $J$ are constants of motion that we call energy and angular momentum. For simplicity of notation we introduce the constants
\begin{equation}
A_\pm \equiv J^2-\LF 1\pm E\RF^2, \qquad B_\pm \equiv \LF J\pm 1\RF^2 - E^2.\label{eq:AB_pm}
\end{equation}
Generic solutions to the geodesic equations are
\begin{eqnarray}
\vp\LF \la\RF &=& \vp_0 + \frac{1}{2 r_H}\ln\LF \frac{e^{\la-\la_0} + B_-e^{-\la+\la_0}}{e^{\la-\la_0} + B_+ e^{-\la+\la_0}}\RF,\label{eq:phi_btz}\\
r^2\LF \la\RF &=& \frac{r_H^2}{4}\LF e^{\la-\la_0} +B_+ e^{-\la+\la_0}\RF\LF e^{\la-\la_0} + B_- e^{-\la+\la_0}\RF,\label{eq:r_btz}\\
v\LF \la\RF &=& v_0 + \frac{1}{2r_H}\ln\left| \frac{ \LF r\LF \la\RF-r_H\RF\LF e^{\la-\la_0} + A_+ e^{-\la+\la_0}\RF}{\LF r\LF \la\RF +r_H\RF \LF e^{\la-\la_0}+ A_-e^{-\la+\la_0}\RF}\right|,\label{eq:v_btz}
\end{eqnarray}
where $\la_0$ is an arbitrary constant that can be viewed as a shift in the affine parameter. The above solutions solve the BTZ equations of motion for any complex $\la_0$. However, we are only interested in solutions for which the coordinates are real, so here we set $\la_0$ to be real and in particular we choose it to be $\la_0=0$.  Boundary anchored geodesics have endpoints at $\la_1 = -\la_2 \rightarrow -\infty$. The two integration constants $\vp_0$, $v_0$ and the constants of motion $E$, $J$ are easily obtained by demanding that the geodesic has endpoints on the boundary at $v\LF \lambda_{1}\RF= t^{(1)}$, $v\LF \lambda_{2}\RF = t^{(2)}$, $\vp\LF \lambda_1\RF=\vp^{(1)}$ and $\vp\LF \lambda_{2}\RF = \vp^{(2)}$ with $t$ the global boundary time. The integration constants are fixed by
\begin{equation}
t^{(1)} = v_0 + \frac{1}{2r_H}\ln\left| \frac{A_+}{A_-}\right|, \qquad t^{(2)}= v_0, \qquad \vp^{(1)} = \vp_0 + \frac{1}{2r_H}\ln\LF \frac{B_-}{B_+}\RF, \qquad \vp^{(2)}= \vp_0.
\label{eq:btz_endpoints}
\end{equation}
For given boundary endpoints the geodesic is completely fixed. This fixes the constants of motion $E$ and $J$ in terms of the boundary coordinates in the eternal BTZ spacetime.  We will later see that $E$ and $J$ are different in the presence of a shell. In section \ref{sec:EE_BTZ} we will derive the holographic entanglement entropy in the static BTZ background.

\subsection{Spatial geodesics in \ads}\label{sec:adsgeo}

Geodesics in global \ads in infalling coordinates
\begin{equation}
ds^2 = -\LF r^2+1\RF dv^2 + 2dvdr + r^2d\vp^2,
\end{equation}
can be computed similarly to the ones in BTZ by integrating the geodesic equations of motion, but it is even simpler to recognize \ads as an analytic continuation of the BTZ geometry in the complex $r_H$-plane. \ads is retrieved at $r_H=-i$ and analogously a continuation of the constants of motion $E\equiv iE_{\text{AdS}}$ and $J\equiv iJ_{\text{AdS}}$ is needed to have real constants of motion in the \ads geometry. The equations of motion \eqref{eq:EOM_phi}-\eqref{eq:EOM_v} are analytically continued to
\begin{align}
\dot{\vp} &= \frac{\jads}{r^2},\label{eq:EOM_ads1}\\
\dot{r}^2 &= \eads ^2 + \LF r^2 + 1 \RF \LF 1- \frac{\jads^2}{r^2}\RF,\label{eq:EOM_ads2}\\
\dot{v} &= \frac{\eads +\dot{r}}{r^2+1}.\label{eq:EOM_ads3}
\end{align}
The geodesic solutions are then analytical continuations of \eqref{eq:phi_btz}-\eqref{eq:v_btz}. Again we are only interested in real solutions. In particular, the radial coordinate $r$ should be real after analytic continuation. This is only possible if we set 
\begin{equation}
\lambda_0 = \pm \frac{\pi i}{2},
\end{equation}
for which the geodesic solutions are
\begin{eqnarray}
\vp\LF \la \RF &=& \vp_0 -\atan\LF \frac{2\jads}{e^{2\la}-1+\jads ^2-\eads ^2}\RF\nonumber\\
&=&\vp_0 - \text{sgn}(\jads)\arccos\LF\frac{e^{2\lambda}-1+\jads^2-\eads^2}{\sqrt{(e^{2\lambda}-1+\jads^2-\eads^2)^2+4\jads^2}}\RF \nonumber,\\
r^2\LF \la\RF &=& \frac{1}{4}\LT e^{2\la} +2\LF \jads ^2 -\eads ^2 -1\RF + \LF \LF 1-\jads ^2 +\eads ^2\RF^2 + 4\jads ^2\RF e^{-2\la}\RT\nonumber,\\
v\LF \la\RF &=& v_0 + \atan\LF r\RF - \frac{\pi}{2}- \atan\LF \frac{2\eads}{e^{2\la} - \eads ^2 + \jads ^2 +1}\RF\nonumber\\
&=& v_0 + \atan\LF r\RF - \frac{\pi}{2}\nonumber\\
& & -\text{sgn}(\eads)\arccos\LF\frac{e^{2\lambda}+1+\jads^2-\eads^2}{\sqrt{(e^{2\lambda}+1+\jads^2-\eads^2)^2+4\eads^2}}\RF.\label{eq:ads_sol}
\end{eqnarray}
The formulae using $\arctan$ are only valid for $-\pi/2<\varphi<\pi/2$ while the formulae using $\arccos$ are valid for  $-\pi<\varphi<\pi$. The integration constants $v_0$, $\vp_0$ and constants of motion $\eads$,  $\jads$ are again fixed by demanding that the geodesics have endpoints on the boundary at $v(\la\rightarrow +\infty)=t^{(2)}$, $v(\la\rightarrow -\infty)=t^{(1)}$, $\vp(\la\rightarrow +\infty)=\vp^{(2)}$ and $\vp(\la\rightarrow - \infty)=\vp^{(1)}$. We thus have that
\begin{align}
 \vp^{(1)} = \vp_0 - \atan\LF \frac{2\jads}{\jads^2 - 1- \eads^2}\RF, &\quad t^{(1)} = v_0 -\atan\LF \frac{2\eads}{\jads^2 +1 -\eads^2}\RF,\nonumber\\
\vp^{(2)} =\vp_0 &\quad \text{and}\quad t^{(2)} = v_0.\label{eq:ads_boundary}
\end{align}
We will study the length of these geodesics in section \ref{sec:EE_ads}.

\subsection{Holographic entanglement entropies}
In this section we will use the formulae of section \ref{sec:btzgeo} and \ref{sec:adsgeo} to compute the holographic entanglement entropy in static global BTZ and in global \ads. We will then also review how geodesics are computed in a dynamical spacetime, namely the homogeneous thin-shell Vaidya spacetime. We will not present a detailed derivation of geodesics in global Vaidya but rather present a qualitative discussion and refer to \cite{Hubeny:2013dea,Ziogas:2015aja} for a detailed computation. The global Vaidya serves as the zeroth order background of a perturbatively inhomogeneous shell geometry, so we will include the formulae that are relevant to us in section \ref{sec:EE_vaidya}.
\subsubsection{Entanglement entropy in BTZ}\label{sec:EE_BTZ}
Because the asymptotic boundary is located at $r\rightarrow + \infty$, geodesic lengths are infinite and need to be regularized. We introduce an IR bulk cutoff $r=r_c$, assume that the boundary points are reached at the cutoff surface and simply by inverting \eqref{eq:r_btz} and evaluating it on the cutoff surface we arrive at
\begin{equation}
\Lcal = \la_2-\la_1 = \ln\LF \frac{4r_c^2}{r_H^2 \sqrt{ \LF -1+E^2-J^2\RF^2 - 4J^2}}\RF.\label{eq:l_btz1}
\end{equation}
The holographic entanglement entropy follows by applying the HRT formula \eqref{eq:RT}. When restricting to a static interval at time $t$ with endpoints at $\vp^{(1)}=\zeta$ and $\vp^{(2)}=\zeta + \ell$ the energy $E$ vanishes and from \eqref{eq:btz_endpoints} we find that
\begin{equation}
J = \frac{1}{\tanh\LF \frac{r_H\ell}{2}\RF}.\label{eq:J_btz}
\end{equation}
The holographic entanglement entropy consequently reduces to
\begin{equation}
S = \frac{\Lcal}{4G_N}=\frac{1}{2G_N}\ln\LF \frac{2r_c}{r_H}\sinh\LF \frac{r_H\ell}{2}\RF\RF.\label{eq:s_btz}
\end{equation}
The entanglement entropy scales linearly with $c\equiv 3/(2G_N)$ and is cutoff dependent, so formally it diverges as the IR cutoff is taken to infinity. Note that the entanglement entropy only depends on $\ell$ but not on the endpoints separately. The holographic entanglement entropy is constant through time and although we are working with spatially periodic BTZ, the entropy agrees with that in a thermal CFT on a line \cite{Calabrese:2004eu}.
\subsubsection{Entanglement entropy in \ads}\label{sec:EE_ads}
Analogously to the BTZ case, one needs to introduce a fixed cutoff radius $r=r_c$ to regulate the geodesic length in \ads. By inverting the radial component of \eqref{eq:ads_sol} and evaluating it on the cutoff surface, the geodesic length is found to be
\begin{equation}
\Lcal = \ln\LF \frac{4r_c^2}{\sqrt{ 4\jads ^2 + \LF 1-\jads ^2 + \eads ^2\RF^2}}\RF.
\end{equation}
In the special case of an equal time interval on the boundary at $\vp^{(1)}=\zeta$ and $\vp^{(2)}=\zeta +\ell$, the corresponding geodesic has $\eads=0$. The complete geodesic lies on a fixed timeslice and from \eqref{eq:ads_boundary} we find that its angular momentum is
\begin{equation}
\jads = \frac{1}{\tan\LF \frac{\ell}{2}\RF}.
\end{equation} 
Correspondingly the holographic entanglement entropy reduces to the well-known 
\begin{equation}
S =\frac{\Lcal}{4G_N}= \frac{1}{2G_N}\ln\LF 2r_c \sin\LF \frac{\ell}{2}\RF\RF,
\label{eq:s_ads}
\end{equation}
which is the entanglement entropy of a CFT vacuum state \cite{Holzhey:1994we,Calabrese:2004eu,Calabrese:2009qy}.
\subsubsection{Entanglement entropy in \ads-Vaidya}\label{sec:EE_vaidya}
The geometries that have been constructed are static and dual to the vacuum (AdS) and thermal (BTZ) state of the dual CFT. These are both equilibrium states. In this paper we are mainly interested in studying thermalization, an inherently non-equilibrium process.\\

A holographic model that captures a surprisingly large number of phenomena of thermalization is the thin shell Vaidya model~\cite{Balasubramanian:2011ur,Hubeny:2013dea,Ziogas:2015aja,AbajoArrastia:2010yt}. The idea is to model thermalization holographically by the infall of an infinitely thin shell of null dust. As the shell falls deeper into the bulk, the geometry becomes that of a black hole. In fact, the Vaidya geometry is that of a BTZ black hole spacetime at $v>0$ glued to vacuum \ads for $v<0$ along the $v=0$ lightlike surface. The metric is
\begin{equation}
ds^2 = -g(r) dv^2 + 2dvdr + r^2 d\vp^2,
\end{equation}
with 
\begin{equation}
g(r) = 
\left\{
\begin{matrix}
r^2 +1 & \text{when} & v<0,\\
r^2-r_H^2 & \text{when} & v>0.
\end{matrix}
\right.
\end{equation}
We will restrict attention to global Vaidya with periodic identifications $\vp\sim\vp +2\pi$. It models a homogeneous infalling shell. Local observables such as the stress-energy tensor are known to instantly thermalize in the Vaidya quench, since they take vacuum expectation values for $t<0$ and thermal expectation values for $t>0$ (where $t$ is the time that corresponds with $v$ on the boundary). Correlation functions and entanglement entropies on the other hand do not thermalize instantly, because they are not local\footnote{This is in the sense that they depend on multiple boundary points.} and are holographically modeled by extended geometric objects like geodesics. Three different cases can be distinguished. Either the geodesic lies completely in the AdS part of the geometry and the dual correlation function or entanglement entropy will be that of the vacuum. This is the case for boundary times $t<0$, so before the shell has been injected. Secondly, it could lie entirely in the BTZ spacetime in which case the dual quantity takes its thermal value. This happens for times $t \geq \ell/2$ for boundary intervals of size $\ell$. Thirdly, the geodesic can cross the shell and have parts in both patches of the spacetime. It will have two legs in the BTZ spacetime and a leg in the \ads $ $ part of the spacetime that connects the two legs in BTZ. This is the regime that has been associated with thermalization in the CFT and occurs while $0\leq t\leq \ell/2$. We review the construction of geodesics that have endpoints at equal times on the boundary. We use the homogeneous thin shell Vaidya as a zeroth order approximation when the shell is only perturbatively inhomogeneous (see section \ref{sec:shell_perturb}).\\

The geodesic action in Vaidya is given by

\begin{align}
\mathcal{I} =&\int\limits_{\la_1}^{\la_2} d\la \sqrt{ -g(r)\dv^2 + 2\dv\dr +r^2\dvp^2}\nonumber\\
=& \int\limits_{\la_{1}}^{\la_{s_1}} d\la \sqrt{ -(r^2-r_H^2)\dv^2 + 2\dv\dr +r^2\dvp^2}+\int\limits_{\la_{s_1}}^{\la_{s_2}} d\la \sqrt{ -(r^2+1)\dv^2 + 2\dv\dr +r^2\dvp^2}\nonumber\\
&+\int\limits_{\la_{s_2}}^{\la_2} d\la \sqrt{ -(r^2-r_H^2)\dv^2 + 2\dv\dr +r^2\dvp^2},
\end{align}
where we have explicitly split the action into the action of the three legs separately and where $\la_{s_1}$ and $\la_{s_2}$ denote the values of the affine parameter where the geodesic crosses the shell. When $v<0$ (i.e.~$\la_{s_1}\leq \la\leq \la_{s_2}$) the equations of motion are those of AdS and correspondingly the geodesic satisfies~(\ref{eq:ads_sol}). When $v>0$  (i.e.~$\la_{s_1}\geq \la$ or $ \la_{s_2}\leq \la$) the geodesic solutions are those of the BTZ patch~(\ref{eq:phi_btz}-\ref{eq:v_btz}). In total there are 12 integration constants of which 4 are fixed by the boundary conditions. These are $\vp\LF \la\rightarrow \pm \infty\RF=\pm\ell/2$ and $v\LF \la\rightarrow\pm\infty\RF = t$ for equal time intervals and by rotational symmetry the turning point of the geodesic can be located at $\vp=0$. At each shell crossing point there are another 4 constraints to determine the remaining integration constants. Two of the constraints just express continuity across the shell. We introduce the notation $r=r_s$ and $\vp=\vp_s$ for the second crossing point. The other two are refraction conditions derived by demanding extremality of the geodesic in the full spacetime, not just in the two patches separately. To ensure this, we demand that the variation of the action under angular and radial variations on the $v=0$ surface vanishes, which results in the following refraction conditions that are also derived in appendix \ref{app:inhom_refrac}:
\begin{align}
\dv_{\text{BTZ}} &= \dv_{\text{\ads}},\\
\dvp_{\text{BTZ}} &= \dvp_{\text{\ads}}.
\label{eq:refrac_hom}
\end{align}
Continuity of $\dv$ turns into a condition relating $E$ to $\eads$ and the angular component turns into continuity of $J$. Unlike static BTZ or \ads, the integration constants cannot be explicitly solved for in terms of the boundary coordinates, but they can be parametrized as a function of the shell crossing radius $r_s$ which can then be found numerically in terms of the boundary coordinates. We will briefly outline the computation here, but refer to e.g.,~\cite{Ziogas:2015aja} for a detailed computation. Note that by symmetry, it is enough to consider one crossing point, and we denote the integration constants on the second BTZ piece (which is attached to the boundary points $(v=t,\vp=\ell/2)$) by $E$, $J$, $v_0$ and $\vp_0$, and the integration constants for the \ads piece by $\eads$, $\jads$, $v_{0,\text{AdS}}$ and $\vp_{0,\text{AdS}}$. The integration constants $v_0$ and $\vp_0$ can be immediately fixed to $v_0=t$ and $\vp_0=\ell/2$ and by reflection symmetry around the midpoint we have $\eads=0$. Moreover, from the angular equation in \eqref{eq:refrac_hom} and the definitions of $J$ and $\jads$ in \eqref{eq:EOM_phi} and \eqref{eq:EOM_ads1} it follows that $r_H J=\jads$. Note however that $E\neq0$, and thus the geodesic does not lie on a constant time slice. From the geodesic solutions in the \ads $ $part \eqref{eq:ads_sol}, one can determine the geodesic length of the part in \ads, and the angle where it crosses the shell $\vp_s$ in terms of $\jads$ and the shell crossing radius $r_s$. The refraction condition for $\dot{v}$ can now be used to write $E$ as a function of $J$ and $r_s$ (see \cite{Ziogas:2015aja}), and the result is
\begin{equation}
r_H E=-\frac{r_H^2+1}{2r_s}\sqrt{\frac{r_s^2-r_H^2J^2}{r_s^2+1}}.\label{eq:E_of_J}
\end{equation}
By evaluating the \ads geodesic solutions on the shell, one can also determine $\dot{v}$ and $\vp_s$ on the shell as a function of $r_s$ and $J$. At the second crossing point we have
\begin{align}
\dot{v} &= \frac{1}{r_s}\sqrt{\frac{r_s^2-r_H^2J^2}{r_s^2+1}},\label{eq:vdot_shell_vaidya}\\
\vp_s &=  \arctan\LF \frac{1}{J}\sqrt{\frac{r_s^2-r_H^2J^2}{r_s^2+1}}\RF,\label{eq:phi_shell_vaidya}
\end{align}
and at the first crossing point they would have the opposite sign. Equation \eqref{eq:vdot_shell_vaidya} is easy to obtain from \eqref{eq:EOM_ads2} and \eqref{eq:EOM_ads3} using $\eads=0$. Equation \eqref{eq:phi_shell_vaidya} is obtained by first determining the radius $r_{\text{min}}$ at the turning point, where $\dot{v}=0$, by using \eqref{eq:EOM_ads2} and \eqref{eq:EOM_ads3}, and we have $r_{\text{min}}=\jads$. From \eqref{eq:ads_sol} we then obtain $e^{2\lambda_{\text{min}}}=1+\jads^2$, and by imposing that $\vp=0$ at $\lambda=\lambda_{\text{min}}$, we obtain $\vp_{0,\text{AdS}}=\arctan(1/\jads)$. Now we can use \eqref{eq:ads_sol} to solve for $\lambda$ at the shell where $r=r_s$, and then plugging this value into the expression for $\vp$ in \eqref{eq:ads_sol} gives \eqref{eq:phi_shell_vaidya}.\\

We can now look at the geodesic in the BTZ part. By setting $r(\la_{s_2})=r_s$ and using \eqref{eq:r_btz} we can solve for $\la_{s_2}$ in terms of $E$, $J$ and $r_s$. After that, we can use equation \eqref{eq:E_of_J} and $v(\la_{s_2})=0$ (where $v(\lambda)$ is given by \eqref{eq:v_btz}) to solve for $E$ and $J$ as a function of $r_s$. The calculations are quite involved, and we refer to \cite{Ziogas:2015aja} for details, but the end result is
\begin{align}
r_H E &= \frac{\LF r_H^2+1\RF\LF r_H -r_s \Tcal\RF}{\Tcal\LF 2r_s^2+r_H^2+1\RF - 2r_Hr_s},\\
r_H J &= \frac{r_s \sqrt{ \Tcal^2 r_H^4 + 4\Tcal^2r_H^2 r_s^2 - 4\Tcal r_H^3r_s + 2\Tcal^2r_H^2 + 4\Tcal r_Hr_s + \Tcal^2 - 4r_H^2}}{\Tcal r_H^2 + 2\Tcal r_s^2 - 2r_Hr_s \Tcal},
\label{eq:Jcircle}
\end{align}
with $\Tcal\equiv\tanh\LF r_H t\RF$. The angular geodesic solution with the above $E$ and $J$ evaluated at the shell, which should be equal to \eqref{eq:phi_shell_vaidya}, then implies a constraint from which one would like to solve $r_s$. The constraint is \cite{Ziogas:2015aja}:
\begin{equation}
\frac{\ell}{2} = \atan\LT \frac{2\LF r_H-r_s\Tcal\RF}{\Scal}\RT + \frac{1}{r_H}\atanh\LT \frac{\Scal}{1-r_H^2+2r_Hr_s\Tcal}\RT,\label{eq:phi_vaidya}
\end{equation}
where
\begin{equation}
\Scal = \sqrt{ -4r_H^2 - 4r_H\LF r_H^2-1\RF r_s \Tcal + \LF 1 + r_H^4  + r_H^2\LF 2+4r_s^2\RF\RF \Tcal^2}.
\end{equation}
Unfortunately, we cannot solve \eqref{eq:phi_vaidya} analytically for $r_s$ as a function of $\ell$ and $t$. So one should solve it numerically. Finally, by inverting the radial geodesic solution evaluated at the boundary, one finds the geodesic length of a geodesic that crosses the shell in Vaidya~\cite{Ziogas:2015aja}:
\begin{align}
\Lcal =& 2\ln\LF 2r_c\RF + 2\ln \LF \frac{ \LF r_H^2+1\RF \Tcal + 2r_s\LF r_s \Tcal-r_H\RF}{r_H \LF r_H^2+1\RF \sqrt{1-\Tcal^2}}\RF\nonumber\\
& + \ln \LF \frac{\LF r_H^2+1\RF \Tcal}{\LF r_H^2+1\RF \Tcal + 4r_s\LF r_s \Tcal-r_H\RF}\RF.
\end{align}
Entanglement entropy in homogeneous Vaidya starts out at its vacuum value \eqref{eq:s_ads}, grows because of the homogeneous quench and at $t=\ell/2$ saturates at its thermal value \eqref{eq:s_btz}. This behaviour is often interpreted as a sign of thermalization in the CFT.\\

For a fixed opening interval on the boundary, we wish to follow the time evolution of a geodesic. At $t<0$ the geodesic is completely inside \ads. As soon as $t>0$ it starts crossing the shell. From $t=\ell/2$ onwards the geodesic lies completely inside the BTZ part of the spacetime. For early times a unique geodesic is associated to each boundary interval. At a particular time 
\begin{equation}
t_{\text{cusp}} =\frac{1}{r_H}\atanh\LF \frac{r_H\sqrt{r_H^2+2}}{r_H^2+1}\RF \label{eq:tcusp}
\end{equation}
the $\ell=\pi$ geodesic acquires angular momentum and a new branch of solutions emerges~\cite{Ziogas:2015aja}. Extra long geodesics split off from this branch and eventually saturate, rendering multiple solutions that wind the black hole. The infinite number of winding geodesics that are known to exist in static BTZ are formed dynamically in Vaidya.

\subsection{Differential entropy}\label{sec:review_differential_entropy}
A second measure of correlation in field theory that we will study in this paper is the differential entropy introduced in~\cite{Balasubramanian:2013lsa}. It measures the residual entropy of a family of observers and is specifically interesting from the point of view of bulk reconstruction. In holographic systems it namely computes the length of a closed bulk curve, a feature that we review in the \ads $ $geometry in section \ref{sec:diff_ent_ads} and in the BTZ geometry in section \ref{sec:diff_ent_btz}.\\

Imagine a set of $2K$ equally spaced observers that each have access to an interval $I_j$. The residual entropy $\Ecal$ of this family of observers is
\begin{equation}
\Ecal = \sum\limits_{j=1}^{2K} \LT S\LF I_j\RF - S\LF I_j\cap I_{j+1}\RF\RT.
\end{equation}
Suppose the state is rotationally symmetric and each interval is centered at an angle $\zeta$ on the boundary and has width $\ell(\zeta)$. If the intervals are shifted an amount $\Delta \zeta$ compared to one another, then the residual entropy of the family of observers is
\begin{equation}
\Ecal = \sum\limits_{\zeta} \LT S\LF \ell(\zeta)\RF - S\LF \ell(\zeta)-\Delta\zeta\RF\RT.
\end{equation}
In the continuum limit, this boils down to
\begin{equation}
\Ecal = \int\limits_0^{2\pi} d\zeta \left.\frac{dS(\alpha)}{d\alpha}\right|_{\alpha=\ell(\zeta)},
\end{equation}
which explains the terminology `differential entropy' for $\Ecal$.\\

The concept of differential entropy can be generalized to non-equal time intervals and non-uniform families \cite{Headrick:2014eia}. Start from a continuous family of observers parametrized by a parameter $\zeta$, which could be a general parameter but we will take it to be the boundary angle. Each observer has access to an interval\footnote{In fact each observer has access to a causal diamond $D[A_\zeta]$, which is the region in the boundary through which every inextendible causal curve that intersects it, necessarily intersects the interval $A_\zeta$ as well. This is because the algebra of observables on $A_\zeta$ should be Lorentz invariant and thus naturally associated to $D[A_\zeta]$.} $A_\zeta$ with coordinates of the left endpoint $\gamma_L\LF \zeta\RF$ and coordinates of the right endpoint $\gamma_R\LF \zeta\RF$. The differential entropy of this family is defined as
\begin{equation}
\Ecal \equiv \oint d\zeta \left.\frac{ \partial S\LF \gamma_L\LF \zeta\RF, \gamma_R\LF \zeta^{'}\RF\RF}{\partial \zeta^{'}}\right|_{\zeta^{'}=\zeta}.
\label{eq:Ecal}
\end{equation}
Now assume that every observer has an entropy measured by the HRT formula, so there exists a minimal curve $\Gamma^\mu \LF \la,\zeta\RF$ with endpoints $\gamma_L\LF \zeta\RF$ and $\gamma_R\LF \zeta\RF$ on the boundary. Here $\Gamma^\mu$ parametrizes the coordinates along the curve. The authors of~\cite{Headrick:2014eia} have shown that the differential entropy then reduces to
\begin{align}
\Ecal = \frac{1}{4G_N}\oint d\zeta \Gamma^{'\mu}(\la_B,\zeta) p_\mu (\la_B,\zeta),
\label{eq:Ecal_Myers}
\end{align}
with $p_\mu$ the canonically conjugate momentum to $\Gamma^\mu$, $\la_B$ a value of the affine parameter along the HRT surface in the bulk and where $'$ stands for the derivative with respect to $\zeta$. By using the definition of canonical momentum in a Lagrangian system where the Lagrangian is the length functional of a geodesic, the differential entropy is shown to equal the length of a closed curve~\cite{Balasubramanian:2013lsa,Headrick:2014eia}. 
By considering different families of observers, different curves can be reproduced. When all closed curves of a given geometry can be reproduced as a differential entropy, one could say the geometry has completely been reconstructed from entanglement entropy. This would add evidence for the current credo that holographic geometry might emerge from entanglement in the boundary. As we review here, this seems to work in \ads  but in BTZ there exists a finite region outside the horizon wherein lengths of closed curves cannot be considered as a differential entropy. This region is dubbed the \textit{entanglement shadow}.
\subsubsection{Differential entropy in \ads}\label{sec:diff_ent_ads}
Consider a static family of observers in \ads with $\gamma_L\LF \zeta\RF = \LF t, \zeta\RF$ and $\gamma_R\LF \zeta\RF=\LF t,\zeta + \ell\RF$ for $\zeta \in [0,2\pi[$ and $\ell$ fixed. The holographic entanglement entropy is given by~(\ref{eq:s_ads}), so by (\ref{eq:Ecal}) the corresponding differential entropy equals
\begin{equation}
\Ecal = \int\limits_{0}^{2\pi} d\zeta \frac{\partial S}{\partial \ell} = \frac{2\pi}{4G_N \tan\LF \frac{\ell}{2}\RF}=\frac{2\pi \jads}{4G_N}.\label{eq:diff_ent_ads_formula}
\end{equation}
There exists a shortcut to see that differential entropy has to be proportional to the angular momentum. One could namely evaluate the angular equation of motion \eqref{eq:EOM_ads1} at the turning point at $r=\jads$ where clearly $\dot{\vp}=\LF d\la/d\vp\RF^{-1}=\jads ^{-1}$. We take the affine parameter the proper length so at the turning point $\la =\mathcal{L}/2$. Similarly by rotational symmetry, the turning point should be located at $\vp=\ell/2$. Because of this symmetry, a change in length due to moving the boundary angle should be equivalent to the change in length due to moving the angular location of the turning point. This implies that $d\la / d\vp = d\mathcal{L}/d\ell= \jads$ which proves that differential entropy is proportional to angular momentum. By the same arguments one can also see this from \eqref{eq:Ecal_Myers}. Equation \ref{eq:diff_ent_ads_formula} shows that the differential entropy computes the length of a closed bulk curve, which in this case is a circle at radius $r=\jads$.\\

By continuously changing $\ell$, circles of all circumferences are reproduced. Specifically at $\ell=\pi$ the size of the circle shrinks to a point at the origin of \ads. The complete timeslice and by time translation invariance all of the geometry can be reconstructed from differential entropy, hence there is no entanglement shadow in \ads. 

\subsubsection{Differential entropy in BTZ}\label{sec:diff_ent_btz}
Differential entropy in BTZ is analyzed in a similar fashion. Consider again a static equal time family of observers doing measurements at time $t$ on an interval of size $\ell$ with one endpoint at $(t,\zeta)$ and the other at $(t,\zeta+\ell)$. Geodesics with endpoints separated an angular distance $\ell$ at time $t$, have $E=0$ and angular momentum given by \eqref{eq:J_btz}. Evaluating~(\ref{eq:l_btz1}) for such geodesics leads to 
\begin{equation}
\Ecal = \frac{2\pi r_H}{4G_N \tanh\LF \frac{r_H \ell}{2}\RF} = \frac{2\pi r_H J}{4G_N}.
\label{eq:diff_ent_btz}
\end{equation}
The same shortcut as in \ads can be applied to see that differential entropy computes the circumference of a circle of radius $r_H J$ in BTZ. While $\Ecal$ can become arbitrarily big, it cannot reproduce all circles. Due to the non-trivial homology constraint in BTZ, the HRT formula only computes entanglement entropy on $A$ via a connected geodesic anchored on $\partial A$ as long as $\ell < \ell_{\text{crit}}$ with~\cite{Hubeny:2013gta}
\begin{equation}
\tanh\LF \frac{r_H \ell_{\text{crit}}}{2}\RF = \frac{ \tanh\LF \pi r_H\RF}{2-\tanh\LF \pi r_H\RF}.
\end{equation}
For $\ell> \ell_{\text{crit}}$ the contributing surface is disconnected and consists of the geodesic that wraps the horizon and the geodesic anchored on $\partial A$ but homologous to $\bar{A}$. The critical interval size $\ell_{\text{crit}}$ is always bigger than $\pi$. So the minimal circle that can be reproduced is the one associated to a family of boundary intervals of size $\ell_{\text{crit}}$ and has radius
\begin{equation}
r_{\text{crit}} =r_H \frac{2-\tanh\LF \pi r_H\RF}{\tanh\LF \pi r_H\RF}.
\end{equation}
For all black holes this is bigger than the horizon radius, hence an entanglement shadow exists in BTZ. The entanglement shadow cannot be probed by minimal geodesics but can be probed by winding geodesics, whose length is given by~(\ref{eq:l_btz1}) with an opening size analytically continued to $\ell>2\pi k$ and $k$ the winding number of the geodesic. That winding geodesics probe the entanglement shadow can be immediately inferred from the fact that their turning point lies at $r<r_{\text{crit}}$, namely $r=r_HJ$ where $J$ is given by \eqref{eq:J_btz} with $\ell>2\pi k$. The question whether or not spacelike extremal surfaces connected to the boundary can probe the entire spacetime has also been studied in higher dimensions, see for instance \cite{Nogueira:2013if}.\\

\section{Perturbative analysis}\label{sec:perturbative}
We now turn our attention to the inhomogeneous shell background. When the inhomogeneities in the stress-energy tensor are small, perturbation theory around the homogeneous thin Vaidya shell can be used. In section~\ref{sec:final_perturb} we will first analyze entanglement and differential entropy in the final state assuming that the geometry is \VBTZ  for all times. In section~\ref{sec:shell_perturb} we treat the perturbatively inhomogeneous shell. We model the asymptotic coordinate transformation \eqref{coord_tr} as
\begin{equation}
F\LF \phim \RF = \phim  + \e f\LF \phim \RF,\label{eq:pert_F}
\end{equation}
with $\e$ the small parameter that controls the strength of inhomogeneities. The perturbative boundary stress-energy tensor found from \eqref{stress_energy_CFT_eq} is
\begin{equation}
T \approx \frac{M}{4} + \frac{\e}{2}\LF Mf'-f^{'''}\RF ,
\end{equation}
where the approximation is up to higher orders in $\e$. In contrast to typical hydrodynamical expansions such as~\cite{Balasubramanian:2013oga} we do not assume anything about the scale of spatial variation. Asymptotically close to the boundary, coordinates $(\tp,\rp,\phip)$ are linearly related to $(\tm,\rmin,\phim)$ by
\begin{align}
\rp &\equiv \rmin + \e\delta\rmin + O(\e^2) \approx \rmin - \frac{\e\rmin}{2} \LF f'\LF \phim + \tm\RF + f'\LF \phim - \tm\RF\RF,\label{eq:rp_pert}\\
\phip &\equiv \phim + \e\delta\phim +O(\e^2)\approx \phim + \frac{\e}{2}\LF f\LF\phim+\tm\RF + f\LF \phim - \tm\RF\RF,\label{eq:phip_pert}\\
\tp &\equiv \tm + \e \delta\tm + O(\e^2) \approx \tm + \frac{\e}{2}\LF f\LF \phim + \tm\RF- f\LF \phim-\tm\RF\RF,\label{eq:tp_pert}
\end{align}
where $(\tp,\rp,\phip)$ are the coordinates in which the metric at $v>0$ takes the form \eqref{BTZ_metric} and $(\tm,\rmin,\phim)$ are those in which the metric at $v<0$ is \eqref{eq:ads_metric}. An advantage of the perturbative approach is that we can obtain analytic results for the entanglement entropy and differential entropy. Both in the final state and in the shell background entanglement entropy will be obtained by a straightforward perturbation of the geodesic action.   
\subsection{Final state}\label{sec:final_perturb}
In this section we analytically compute entanglement entropy and differential entropy in the final state \VBTZ $ $ geometry. As it turns out, the first order entanglement entropy is a linear combination of left and right movers and agrees with the CFT result in an inhomogeneous quench \cite{CardySotiriadis2008}. The differential entropy does not receive first order corrections and is time independent.
\subsubsection{Entanglement entropy}

We are interested in the first order perturbed entanglement entropy in \VBTZ which by a coordinate transformation can be brought to the standard BTZ. As such the equal time entropy in \VBTZ could be computed from the length of a non-equal time geodesic in BTZ similar to \eqref{eq:l_btz1}. The energy and angular momentum would again be fixed by the boundary conditions \eqref{eq:btz_endpoints}, but now the boundary coordinates in \eqref{eq:btz_endpoints} would be the $(\bar{t},\bar{\vp})$ coordinates and would thus not be at equal times. Moreover, the cutoff radius would be position dependent. As we have discussed in section \ref{sec:shell}, the choice we make here is that the cutoff is constant in the $(t,r,\vp)$ coordinates and therefore position dependent in the $(\bar{t},\bar{r},\bar{\vp})$ coordinates. We could then perturbatively expand the geodesic length in the inhomogeneities to find the first order perturbed entropy.\\

While the above method is possible in the final state geometry, such a method is much more difficult when a shell is included, because we cannot analytically solve for $E$ and $J$ in terms of the boundary coordinates. For that reason, we apply a different method that can be used both in the final state and in the inhomogeneous shell background. In this section we apply our method to the eternal \VBTZ geometry. We start from the on-shell action
\begin{equation}
\mathcal{I} = \int\limits_{\la_1}^{\la_2} \sqrt{g_{\mu\nu}\dot{x}^\mu \dot{x}^\nu}d\la ,
\end{equation}
where we have defined $\la$ to be the proper length of the unperturbed geodesic. To first order in the inhomogeneities we can write the entanglement entropy in \VBTZ as $S= S^{(0)} + \e\delta S + O(\e^2)$ with $S^{(0)}$ the entropy of a static BTZ and where we have explicitly brought out the small parameter $\e$ from the perturbation. The entropy is related to the on-shell action by $S=\mathcal{I}/4G_N$. In general, the action acquires perturbations both from the metric and from the geodesic solutions $x^{\mu}(\la)$, which are expanded as
\begin{align}
x^{\mu} &= x^{(0)\mu} + \e\delta x^\mu + O(\e^2),\\
g_{\mu\nu}(x) &= g_{\mu\nu}^{(0)}(x) + \e\delta g_{\mu\nu}(x)+O(\e^2)\nonumber\\
&=g_{\mu\nu}^{(0)}\LF x^{(0)}\RF+ \e \partial_\alpha g_{\mu\nu}^{(0)}\LF x^{(0)}\RF\delta x^{\alpha} + \e\delta g_{\mu\nu}\LF x^{(0)}\RF + O(\e^2),
\end{align}
where $x^{(0)\mu}$ and $g^{(0)}_{\mu\nu}$ are the coordinates and metric in the unperturbed static BTZ. By using the lowest order geodesic equations of motion, the first order perturbed action reduces to
\begin{equation}
\delta \mathcal{I} =\frac{1}{2} \int\limits_{\la_1}^{\la_2} \delta g_{\mu\nu} \dot{x}^{(0)\mu} \dot{x}^{(0)\nu} d\la + \LT g^{(0)}_{\mu\nu} \dot{x}^{(0)\mu} \delta x^{\nu}\RT^{\la_2}_{\la_1}.
\end{equation}
This expression is fully covariant and can be evaluated in any coordinate system.
We choose to work in the $(\tp,\rp,\phip)$ coordinate system defined by \eqref{BTZ_metric} where the geometry is again that of a static BTZ. This simplifies $\delta \mathcal{I}$ considerably, because it is the frame in which $\delta g_{\mu\nu}=0$. Instead, the perturbations $\delta x^\mu$ of the boundary endpoints of the geodesic are not zero but are given by equations \eqref{eq:rp_pert}-\eqref{eq:tp_pert}. For ease of notation, we will drop the label `(0)' on the unperturbed quantities whenever we use an explicit coordinate system. The first order entropy becomes a boundary term. The linear relations between the barred and non-barred coordinate sets~\eqref{eq:rp_pert}-\eqref{eq:tp_pert} imply that
\begin{equation}
\delta \mathcal{I}= \LT \frac{ \dot{r} \delta \rmin}{\rmin^2 -M}\RT^{\lambda_2}_{\lambda_1} - \LT \LF \rmin^2 -M\RF \dot{t}\delta\tm\RT^{\lambda_2}_{\lambda_1} + \LT \rmin^2 \dot{\vp} \delta\phim\RT^{\lambda_2}_{\lambda_1}.\label{eq:var_btz}
\end{equation}
Remember that to zeroth order the barred and non-barred coordinates are the same. We will restrict to equal time intervals in the $(t,r,\vp)$ coordinates such that the unperturbed geodesic has $\dot{t}=0$. Likewise the zeroth order BTZ equations of motion tell us that $\rmin^2\dot{\vp} = r_H J$ and 
\begin{equation} 
\LT \frac{\dot{r}\delta r}{r^2-M}\RT^{\la_2}_{\la_1} = \frac{\delta r(\la_2) + \delta r (\la_1)}{r_c} + O\LF \frac{1}{r_c^2}\RF,
\end{equation}
in a near boundary expansion. We use (\ref{eq:rp_pert}) and (\ref{eq:phip_pert}) to express $\delta \phim$ and $\delta\rmin$ in terms of $f$. The first order perturbation of entanglement entropy  of an interval of size $\ell$ starting at an angle $\zeta$ and at time $t$ becomes
\begin{align}
4G_N\delta S &= -\frac{1}{2}\LT f'\LF\zeta + \ell+t\RF + f'\LF\zeta +\ell-t\RF + f'\LF\zeta+t\RF + f'\LF\zeta-t\RF\RT\nonumber\\
& + \frac{r_HJ}{2}\LT f\LF\zeta+\ell+t\RF + f\LF\zeta+\ell-t\RF -  f\LF\zeta+t\RF - f\LF\zeta-t\RF\RT.
\label{eq:delta_S_btz}
\end{align}
It is a simple linear combination of left and right movers. We only needed the zeroth order solutions for an equal time geodesic in the BTZ background and the coordinate transformation from \VBTZ to a static BTZ to first order at the boundary. It can be checked that explicitly solving for the geodesic to first order and computing its length gives the same result. Remarkably the same expression has been found in~\cite{CardySotiriadis2008} for the late time entropy after a perturbative inhomogeneous quench in $d=1+1$ CFT. First order inhomogeneous corrections to entanglement entropy therefore seem to be governed by conformal symmetry. When the perturbation of the action is written using the $(t,r,\vp)$ metric instead, one retrieves a first law like formula for the entanglement entropy, namely as an integral over the stress-energy tensor perturbations, which after performing the integral over $\la$, of course gives the same result as \eqref{eq:delta_S_btz}. For completeness, since we use it in figures \ref{fig_inhom_entropy_2} and \ref{fig_inhom_entropy_2b}, we also quote the second order result here, which is
\begin{align}
4G_N\delta S^{(2)}=&-\frac{r_H^2}{4(\cosh(r_H\ell)-1)}\left((f(-t+\zeta)-f(-t+\zeta+\ell))^2+(f(t+\zeta)-f(t+\zeta+\ell))^2\right)\nonumber\\
&+\frac{1}{4}(f'(-t+\zeta)^2+f'(t+\zeta)^2+f'(-t+\zeta+\ell)^2+f'(t+\zeta+\ell)^2).
\end{align}
This can be obtained by expanding the exact formula \eqref{eq:L_VBTZ}, which is derived in section \ref{sec:non_perturbative}, to second order in $\epsilon$.

\subsubsection{Differential entropy}\label{sec:final_pert_diffent}
We will consider the differential entropy in \VBTZ of a family of intervals of size $\ell$ at time $t$, parametrized by angle $\zeta$ (so that the two enpoints of an interval are given by $(t,\zeta)$ and $(t,\zeta+\ell)$). It can be perturbed to first order as 
\begin{equation}
\Ecal = \Ecal^{(0)} + \e \delta \Ecal,
\label{eq:diff_ent_vbtz}
\end{equation}
with $\Ecal^{(0)}$ the differential entropy of the unperturbed static BTZ given by~(\ref{eq:diff_ent_btz}) and 
\begin{equation}
\delta\Ecal = \oint d\zeta \frac{\partial \delta S}{\partial\ell}\LF \zeta,\zeta+\ell\RF.
\label{eq:diff_ent_vbtz2}
\end{equation}
To compute $\delta\Ecal$ we plug~(\ref{eq:J_btz}) into~(\ref{eq:delta_S_btz}) and differentiate $\delta S$ with respect to $\ell$. We find
\begin{align}
4G_N\frac{\partial \delta S}{\partial \ell} = -\frac{1}{2}&\LT f^{''}\LF \zeta + \ell+t\RF + f^{''}\LF \zeta + \ell -t\RF\RT + \frac{r_H J}{2}\LT f^{'}\LF \zeta + \ell + t\RF + f^{'}\LF \zeta +\ell-t\RF\RT \nonumber\\
&+ \frac{r_H}{2}\frac{dJ}{d\ell}\LT f\LF \zeta + \ell +t\RF + f\LF \zeta+ \ell -t\RF - f\LF \zeta+t\RF -f\LF \zeta-t\RF\RT.
\end{align}
The first two terms in this expression vanish when integrating over $\zeta$ because they only involve total derivatives in $\zeta$ and because $f$ is necessarily periodic with period $2\pi$. Integration over $\zeta$ involves the whole period of $f$. Using this, the four contributions to the last term vanish two by two. So the differential entropy does not receive first order corrections in \VBTZ and equals
\begin{equation}
\Ecal = \frac{2\pi r_H}{4G_N \tanh\LF \frac{r_H \ell}{2}\RF} + O(\e^2).
\end{equation}
\subsection{Inhomogeneous shell}\label{sec:shell_perturb}
When an inhomogeneous shell of massless matter is injected at $t=0$ on the boundary, the geometry is \VBTZ when $v>0$ and \ads when $v<0$. In this section we analyze the entanglement entropy and differential entropy when the shell is only perturbatively inhomogeneous. The procedure that we will follow is completely analogous to section~\ref{sec:final_perturb} but the presence of the shell makes a lot of formulae more tedious.
In this case the entanglement entropy does not agree with \cite{CardySotiriadis2008}, but we do not expect agreement during the quench since the setups are different (we only expect agreement for the final state which is governed by conformal symmetry). In particular this means that the holographic entropy is not reproduced by a quasi particle picture. The differential entropy to lowest order monotonically interpolates between its result in \ads and in BTZ and does not receive first order corrections from the inhomogeneities.
\subsubsection{Entanglement entropy}
As in section~\ref{sec:final_perturb}, a first order correction to the homogeneous entanglement entropy can be analytically obtained from a perturbation of the geodesic action. The procedure is the same, namely we start with the zeroth order solution which is the geodesic in the presence of a homogeneous shell which is constructed in section \ref{sec:EE_vaidya}. We then consider small perturbations of this geodesic such that it solves the equations of motion for a geodesic in the presence of an inhomogeneous shell where the inhomogeneities are small. However, just as in \ref{sec:final_perturb}, in order to obtain the perturbative correction to the action (and thus the entanglement entropy) it is not necessary to obtain the explicit solutions for the geodesic. Note that for boundary times $t<0$ the geodesic is trivially equal to that of \ads, and for $t>\ell/2$ the solutions are those obtained in section \ref{sec:final_perturb}. We will thus here only consider the non-trivial time interval $0<t<\ell/2$, where the geodesic intersects the shell (see appendix \ref{app:saturation}) and thus consists of two pieces in the \VBTZ spacetime and one piece in the \ads spacetime, and these are all glued together across the shell at $v=0$.  The total geodesic action can thus be written as
\begin{equation}
\mathcal{I}=\int\limits_{\la_{s2}}^{\la_2} \sqrt{g^{>}_{\mu\nu}\dot{x}^\mu \dot{x}^\nu}d\la + \int\limits_{\la_1}^{\la_{s1}} \sqrt{g^{>}_{\mu\nu}\dot{x}^\mu \dot{x}^\nu}d\la + \int\limits_{\la_{s1}}^{\la_{s2}} \sqrt{g^{<}_{\mu\nu}\dot{x}^\mu \dot{x}^\nu}d\la ,
\label{eq:geod_action_shell}
\end{equation}
with $\la_{i}$ the boundary values of the affine parameter and  $\la_{s_i}$ the values of the affine parameter where the geodesic hits the shell at $v=0$. The notation `$>$' means above the shell (in the \VBTZ geometry) and `$<$' means below the shell (in the \ads geometry). Perturbations to the action will again arise due to perturbations of either the metric or the geodesic solution. To lowest order the action describes a geodesic in homogeneous global Vaidya described in~\cite{Hubeny:2013dea,Ziogas:2015aja}  and in section \ref{sec:EE_vaidya}. Since all final \VBTZ solutions are related by a coordinate transformation to the static BTZ black hole, we can choose coordinates such that the final spacetime is a BTZ black hole and thus $\delta g_{\mu\nu}=0$. In these coordinates, all effects of the perturbations are coming from the perturbation of the geodesic. Writing the perturbation of the geodesic as $\delta x^\mu$ and the zeroth order solution of homogeneous Vaidya as $x^{(0)\mu}$, the perturbation of the action, after using the equations of motion, reduces to a sum of boundary terms as
\begin{align}
\delta \mathcal{I}= \LT g^{(0)>}_{\mu\nu} \dot{x}^{(0)\mu} \delta x^\nu\RT^{\la_2}_{\la_{s2}} + \LT g^{(0)>}_{\mu\nu} \dot{x}^{(0)\mu} \delta x^{\nu}\RT^{\la_{s1}}_{\la_1}+\LT g^{(0)<}_{\mu\nu} \dot{x}^{(0)\mu} \delta x^\nu\RT^{\la_{s2}}_{\la_{s1}}.\label{eq:deltaI}
\end{align}
From here on we will drop the `$(0)$' labels on the unperturbed quantities, and the unperturbed coordinates will for $v>0$ represent the static BTZ coordinates $(v,\bar{r},\phip)$ and for $v<0$ we choose the \ads coordinates $(v,r,\vp)$.  Note that the tradeoff of choosing coordinates where the final spacetime is a static BTZ black hole is that instead the boundary points of the geodesic are affected by the perturbation, namely $\delta x^\mu(\lambda_i)\neq0$. To obtain the perturbation to the action, we thus need both the perturbations of the geodesic at the boundary and at the shell. At the boundary they are given by~\eqref{eq:rp_pert}-\eqref{eq:tp_pert}, since this is the asymptotic (linearized) coordinate transformation that maps the \VBTZ solution to the static BTZ solution. At the shell, we first have that the shell crossing radius and shell crossing angles will receive perturbative corrections compared to their homogeneous values. On top of that, the barred coordinates are related to the non-barred ones by the coordinate transformation \eqref{coord_tr} on the shell, which, together with \eqref{eq:pert_F}, gives the following relations
\begin{align}
r\LF \la_{si}\RF & = r_s + \e \delta r_{s}^{(i)},\label{eq:shell_pert_1}\\
\rp\LF \la_{si}\RF &= r_s + \e \LF \delta r_{s}^{(i)} -r_s f'\LF \vp_{s}^{(i)}\RF\RF,\\
\vp\LF \la_{si}\RF &= \vp_s^{(i)} + \e \delta \vp_{s}^{(i)},\\
\phip\LF \la_{si}\RF &= \vp_s^{(i)} + \e\LF \delta\vp_{s}^{(i)} + f\LF \vp_{s}^{(i)}\RF \RF\label{eq:shell_pert_4},
\end{align}
where $r_s$ and $\vp_s^{(i)}$ are the homogeneous shell crossing radius and angles and $\delta r_s^{(i)}$ and $\delta \vp_s^{(i)}$ are the perturbations at the crossing points $i=1,2$. The shell is always located at $v=0$ so the $v$-component of the perturbation on the shell vanishes. Using continuity of $\dot{v}$ and $\dot{\vp}$ along the shell given by the homogeneous refraction conditions~(\ref{eq:refrac_hom}), and plugging the perturbations \eqref{eq:shell_pert_1}-\eqref{eq:shell_pert_4} into \eqref{eq:deltaI} yields the contribution to the action from the shell crossing points as
\begin{equation}
\left.\delta \mathcal{I}\right|_{\text{shell}} = r_s\LT \dot{v}_{s}^{(2)} f'\LF \vp^{(2)}_{s}\RF - \dot{v}_{s}^{(1)} f'\LF\vp^{(1)}_{s}\RF\RT - r_H J\LF f\LF \vp^{(2)}_s\RF - f\LF\vp^{(1)}_s\RF\RF.
\end{equation}
Now let us denote the endpoints of the geodesic by $\vp^{(1)}=\zeta$ and $\vp^{(2)}=\zeta+ \ell$. To lowest order the shell crossing angle can then be taken to lie at $\vp_s^{(1)}\equiv \zeta + \ell/2 -\vp_s$ and $\vp_s^{(2)}\equiv\zeta+\ell/2 + \vp_s$, where $\vp_s$ is given by \eqref{eq:phi_shell_vaidya}. The angular momentum $J$ is that of a geodesic in homogeneous Vaidya. From equation \eqref{eq:vdot_shell_vaidya} we know $\dot{v}$ on the shell as a function of $r_s$ at $\la=\la_{s2}$. At $\la=\la_{s1}$ one would have opposite sign.\\

Now let us consider the contributions from the boundary to \eqref{eq:deltaI}. The contribution from the boundary, after changing from the coordinates $(v,\bar{r},\phip)$ to $(\bar{t},\bar{r},\phip)$, is given by
\begin{equation}
\left.\delta \mathcal{I}\right|_{\text{boundary}}= \LT \frac{ \dot{r} \delta \rmin}{\rmin^2 -M}\RT^{\lambda_2}_{\lambda_1} - \LT \LF \rmin^2 -M\RF \dot{t}\delta\tm\RT^{\lambda_2}_{\lambda_1} + \LT \rmin^2 \dot{\vp} \delta\phim\RT^{\lambda_2}_{\lambda_1}.\label{eq:var_btz_2}
\end{equation}
The calculation here is similar to that of section~\ref{sec:final_perturb} but the lowest order geodesic is not that of static BTZ but of homogeneous Vaidya, which means the geodesics are not equal time but have an energy $E$ that can be determined as a function of $r_s$ which is given by \eqref{eq:Jcircle}. The perturbative relation between the barred and unbarred coordinates is given by \eqref{eq:rp_pert}-\eqref{eq:tp_pert} from which we obtain expression for $\delta x^\mu$, and from \eqref{eq:EOM_phi}-\eqref{eq:EOM_v} we have that $(r^2-M)\dot{t}=r_HE$, $r^2\dot{\phim}=r_H J$ and $\dot{r}=\pm r+O(1)$ where the minus (plus) sign corresponds to $\lambda_1$ ($\lambda_2$). Using these results, \eqref{eq:var_btz_2} becomes
\begin{align}
\left.\delta \mathcal{I}\right|_{\text{boundary}} &= -\frac{1}{2}\LT f'(\vp^{(2)}+t) + f'(\vp^{(2)}-t) + f'(\vp^{(1)}+t) + f'(\vp^{(1)}-t)\RT\nonumber\\
& + \frac{r_HJ}{2}\LT f(\vp^{(2)}+t) + f(\vp^{(2)}-t) -  f(\vp^{(1)}+t) - f(\vp^{(1)}-t)\RT \nonumber\\
& + \frac{r_HE}{2} \LT f(\vp^{(2)}-t) + f(\vp^{(1)}-t) - f(\vp^{(1)}+t)-f(\vp^{(2)}+t)\RT.
\end{align}
By adding all contributions we find the total entanglement entropy for the interval $(\vp^{(1)},\vp^{(2)})=(\zeta,\zeta+ \ell)$ to be
\begin{align}
S &= S^{(0)} + \e\left. \delta S\right|_{\text{shell}} + \e\left. \delta S\right|_{\text{boundary}},\nonumber\\
&= S^{(0)}  -\frac{\e}{8G_N}\LT f'(\zeta+\ell+t) + f'(\zeta+\ell-t) + f'(\zeta+t) + f'(\zeta-t)\RT\nonumber\\
& + \e\frac{r_HJ}{8G_N}\LT f(\zeta+\ell+t) + f(\zeta+\ell-t) -  f(\zeta+t) - f(\zeta-t)\RT \nonumber\\
& + \e\frac{r_HE}{8G_N} \LT f(\zeta+\ell-t) + f(\zeta-t) - f(\zeta+t)-f(\zeta+\ell+t)\RT  \nonumber\\
& +\frac{\e}{4G_N}\sqrt{ \frac{r_s^2- r_H^2 J^2}{r_s^2+1}}\LT  f'\LF  \zeta +\frac{\ell}{2}+ \vp_s\RF +  f'\LF\zeta+ \frac{\ell}{2}-\vp_s\RF\RT\nonumber\\
& - \frac{\e r_H J}{4G_N} \LF f\LF \zeta +\frac{\ell}{2}+ \vp_s\RF - f\LF\zeta+ \frac{\ell}{2}-\vp_s\RF\RF + O(\e^2).
\label{eq:S_shell}
\end{align}
The first order entanglement entropy is again a linear combination of left and right movers, up to the non-trivial time dependence of $E,J,r_s$ and $\vp_s$.
\subsubsection{Differential entropy}
The differential entropy can be readily computed from entanglement entropy. In fact the differential entropy has not been considered yet for homogeneous Vaidya. In this section we analyze the differential entropy to first order in $\e$.
\paragraph{Homogeneous Vaidya}\mbox{}\\

\begin{figure}[h]
\centering
\begin{subfigure}{0.4\textwidth}
\centering
\includegraphics[scale=0.3]{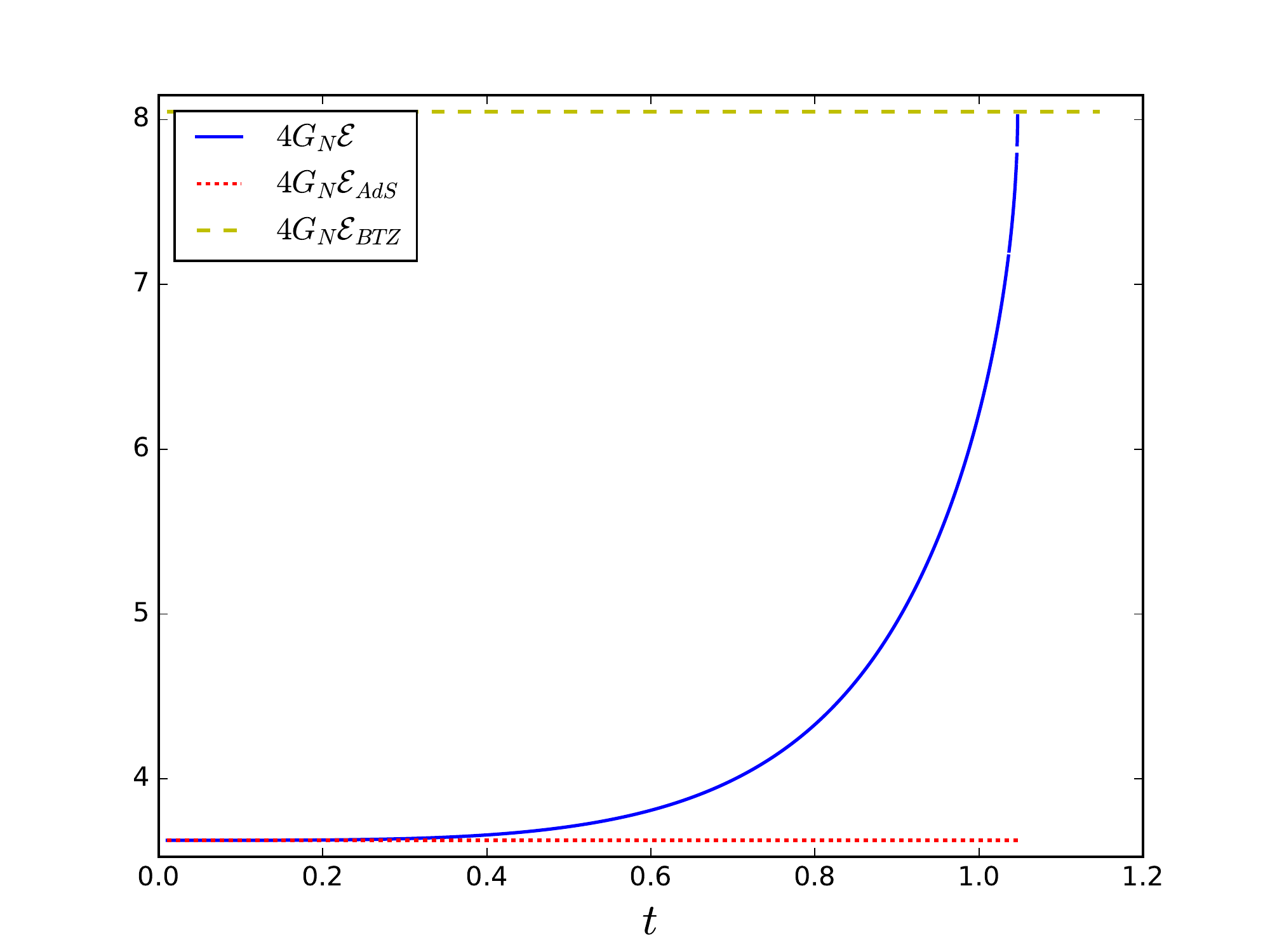}
\caption{}
\label{fig:diff_ent_2PiOver3}
\end{subfigure}\quad
\begin{subfigure}{0.4\textwidth}
\centering
\includegraphics[scale=0.3]{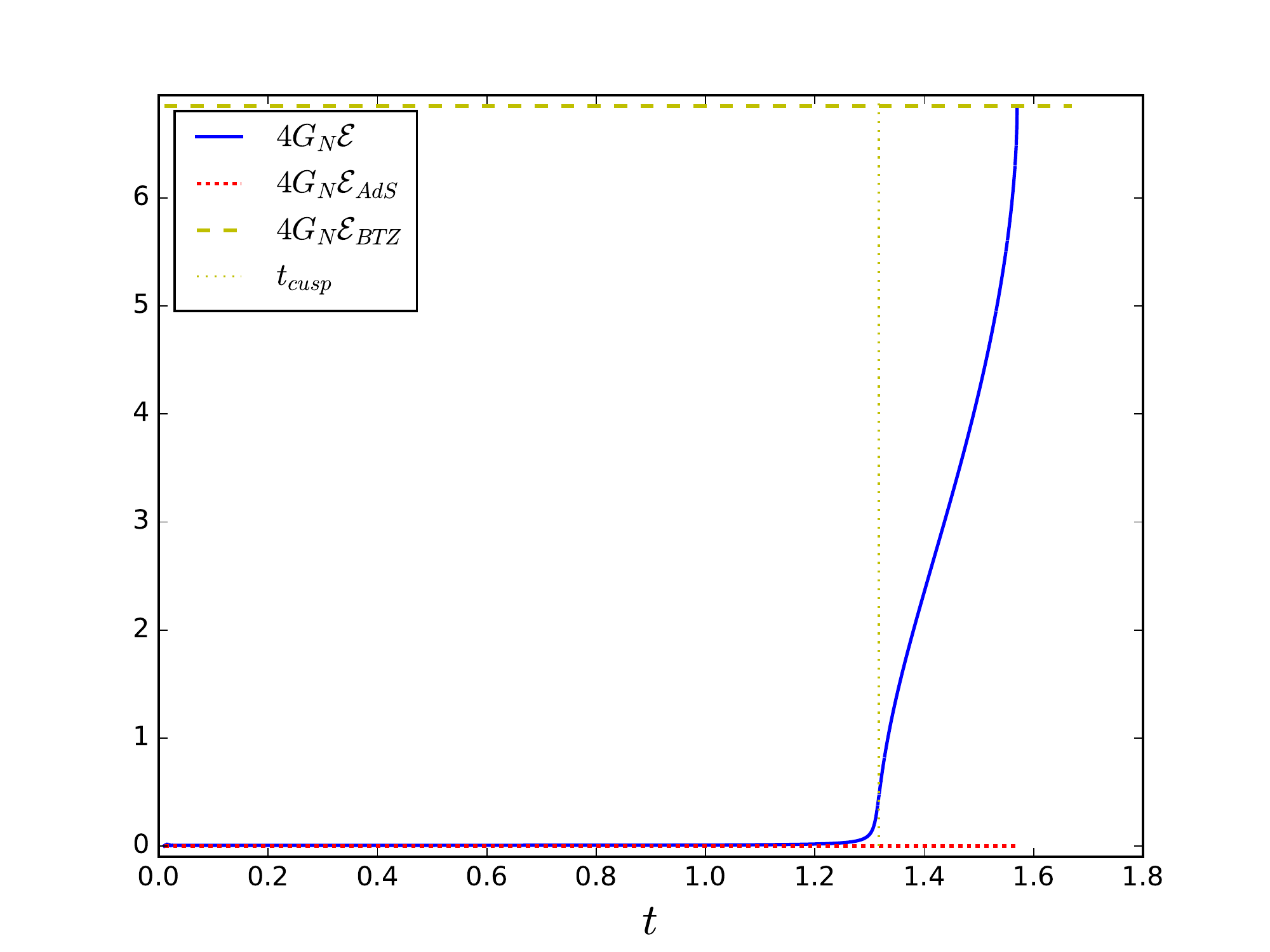}
\caption{}
\label{fig:diff_ent_Pi}
\end{subfigure}
\caption{Differential entropy up to $O(\e^2)$ as a function of time. Figure (a) plots $\Ecal$ for an interval of size $\ell= 2\pi/3$. Figure (b) plots $\Ecal$ for an interval of size $\ell=\pi-0.002$. In both figures $r_H=1$.}
\label{fig:diff_ent_vaidya}
\end{figure}

The differential entropy in a homogeneous Vaidya background can be computed with a similar logic as the shortcut described in section \ref{sec:diff_ent_ads}. This immediately gives
\begin{equation}
\Ecal^{(0)} = \frac{2\pi r_H J}{4G_N},
\end{equation}
where we choose conventions such that $r_H J=\jads$ the angular momentum along the geodesic. In contrast with the static geometries, the angular momentum is time dependent in Vaidya and given by \eqref{eq:Jcircle}. As explained before, we do not have an analytic formula for $r_s$ but can compute it numerically by inverting~\eqref{eq:phi_vaidya}. Once $r_s(t,\ell)$ is known numerically, the differential  entropy can be plotted for any fixed $\ell$ as a function of $t$. As an example, we have plotted its behaviour through time for $\ell=2\pi/3$ in figure~\ref{fig:diff_ent_2PiOver3}. $\Ecal$ starts out at its \ads value and from $t>0$ onwards monotonically grows until it saturates at its BTZ value at $t=\ell/2$. From that time on it will be constant. From figure~\ref{fig:diff_ent_2PiOver3} it would seem the saturation is rather abrupt, but one can show by expanding~(\ref{eq:Jcircle}) in $(\ell/2-t)$ that the transition is smooth.\\ 

For intervals of size $\ell=\pi$ the differential entropy is zero for times smaller than the cusp time \eqref{eq:tcusp}. Before that time the corresponding geodesic has angular momentum zero and goes through the origin at $r=0$. Since this geodesic is the minimal surface that probes deepest into the bulk, the growth of differential entropy from $t_{\text{cusp}}$ onwards signals the dynamical formation of an entanglement shadow in Vaidya. To probe inside the entanglement shadow one needs to resort to the long geodesics that are formed dynamically from $t_{\text{cusp}}$ onwards. Due to numerical difficulties we cannot plot the exact formation of the entanglement shadow but we can plot the differential entropy infinitesimally close to $\ell=\pi$. The result is depicted in figure \ref{fig:diff_ent_Pi}. Because of the trivial homology constraint in Vaidya, the disconnected geodesic configuration does not contribute to entanglement entropy nor to the differential entropy.
 
\paragraph{Perturbatively inhomogeneous shell}\mbox{}\\

Differential entropy in the inhomogeneous shell background can be computed from~(\ref{eq:S_shell}) in a first order approximation in inhomogeneities. We use the definition~(\ref{eq:diff_ent_vbtz}) together with~(\ref{eq:diff_ent_vbtz2}) and compute the derivative of $\delta S$ with respect to $ \ell$. With an argument similar to that of section~\ref{sec:final_pert_diffent} we can show that 
\begin{equation}
\delta \Ecal=0.
\end{equation}
This proves that the differential entropy does not receive first order corrections in the inhomogeneities and equals
\begin{equation}
\Ecal = \frac{2\pi r_H J}{4G_N}+O(\epsilon^2).
\end{equation}

\section{Non-perturbative analysis}\label{sec:non_perturbative}
When the inhomogeneities are non-perturbative it is still possible to treat the entanglement entropy of \VBTZ analytically, although in the presence of the shell analytic formulas are difficult to obtain and the parameters for the geodesics, such as the energies and angular momenta, can not be obtained analytically in terms of the boundary coordinates and must be obtained numerically. In this section we discuss entanglement and differential entropy both in the final state and in the inhomogeneous shell background. We derive refraction conditions on the shell and present the numerical algorithm that we implement to compute the entropy.

\subsection{Final state}\label{sec:final_non_pert}
In this section we obtain analytical results for entanglement entropy and differential entropy. In particular we see that also non-perturbatively the differential entropy is time independent.
\subsubsection{Entanglement entropy}
Recall that \VBTZ can be transformed to static BTZ  in coordinates $(v,\rp,\phip)$. Because the geodesic length is diffeomorphism invariant, the holographic entanglement entropy formula should be given by \eqref{eq:l_btz1} but with position dependent cutoffs, such that 
\begin{equation}
\Lcal =\ln\LF \frac{4r_c^{(1)}r_c^{(2)}}{r_H^2 \sqrt{ \LF -1+E^2-J^2\RF^2 - 4J^2}}\RF\label{eq:l_btz}.
\end{equation} 
The constants of motion $E$ and $J$ are again related to the boundary coordinates by \eqref{eq:btz_endpoints} but in this relation one has to use the barred coordinates $(\bar{t},\bar{\vp})$. Remember that the  endpoints $(\phip^{(1)},\tp^{(1)})$ and $(\phip^{(2)},\tp^{(2)})$ in the static BTZ coordinate system are not at equal times and are related to $(\vp^{(1)},t^{(1)})=(\zeta,t)$ and $(\vp^{(2)},t^{(2)})=(\zeta + \ell, t)$ by 
\begin{align}
\phip^{(1)}=&\frac{1}{2}\LF F(t+\zeta)+F(-t+\zeta) \RF,\quad \tp^{(1)}=\frac{1}{2}\LF F(t+\zeta)-F(-t+\zeta) \RF\label{eq:boundary_points},\\
\phip^{(2)}=&\frac{1}{2}\LF F(t+\zeta+\ell)+F(-t+\zeta+\ell) \RF,\quad \tp^{(2)}=\frac{1}{2}\LF F(t+\zeta+\ell)-F(-t+\zeta+\ell) \RF.\nonumber
\end{align}
So after using the coordinate transformation to the $(t,\vp)$ boundary coordinates (because it is these coordinates in which the interval is at equal times), the energy and angular momentum of the geodesic are constrained by
\begin{equation}
\frac{1}{2r_H}\ln\frac{J^2-(1+ E)^2}{J^2-(1- E)^2}=\frac{1}{2}[F(t+\zeta)-F(-t+\zeta)-F(t+\zeta+\ell)+F(-t+\zeta+\ell)]\equiv a,\label{eq:EJeq1}
\end{equation}
\begin{equation}
\frac{1}{2r_H}\ln\frac{(J-1)^2-E^2}{(J+1)^2-E^2}=\frac{1}{2}[F(t+\zeta)+F(-t+\zeta)-F(t+\zeta+\ell)-F(-t+\zeta+\ell)]\equiv b,\label{eq:EJeq2}
\end{equation}
where we have denoted the right hand side of~\eqref{eq:EJeq1} by $a$ and the right hand side of~\eqref{eq:EJeq2} by $b$ for ease of notation. With a bit of algebra these equations can be solved for $E$ and $J$ as
\begin{align}
E_\pm &= \frac{  e^{2r_H a}-1}{\LF e^{r_H (a-b)} \pm 1\RF\LF e^{r_H (a+b)}\pm 1\RF}=\frac{\sinh(r_H a)}{\cosh(r_H a)\pm\cosh(r_H b)},\label{eq:E_pm}\\
J_\pm &= \sqrt{ \frac{ \LF 1+E_\pm\RF^2 - e^{2r_Ha}\LF 1-E_\pm\RF^2}{1-e^{2r_Ha}}}=\frac{\mp\sinh(r_Hb)}{\cosh(r_Ha)\pm\cosh(r_Hb)}.\label{eq:J_pm}
\end{align} 
By~(\ref{eq:r_cutoff}) the uniform bulk IR cutoff $r_c$ is related to the position dependent cutoffs $r_c^{(1)}$ and $r_c^{(2)}$. When we plug them into~(\ref{eq:l_btz}) the geodesic length in \VBTZ is fixed to
\begin{equation}
\Lcal=\ln\left|\frac{4r_c^2 \LF \cosh(r_H a) \pm \cosh(r_Hb)\RF}{r_H^2\sqrt{F'(\zeta+t)F'(\zeta-t)F'(\zeta+\ell+t)F'(\zeta+\ell-t)}}\right|\label{eq:L_VBTZ}.
\end{equation}
By demanding that we retrieve the integration constants of static BTZ when $F(\vp)=\vp$ it becomes clear that we need to use the lower sign in the above expressions.
\subsubsection{Differential entropy}
From the entanglement entropy one can derive a formula for differential entropy from the usual definition
\begin{equation}
\Ecal = \int\limits_{0}^{2\pi} d\zeta \frac{\partial S}{\partial \ell}.
\end{equation}
The entanglement entropy of \VBTZ is given by~\eqref{eq:L_VBTZ}, which together with~\eqref{eq:E_pm} and~\eqref{eq:J_pm} results in
\begin{align}
4G_N\frac{\partial S}{\partial\ell} = -\frac{1}{2} &\LT \frac{F^{''}\LF \zeta+\ell +t\RF}{F'\LF \zeta +\ell +t\RF} + \frac{F^{''}\LF \zeta+\ell -t\RF}{F'\LF \zeta +\ell -t\RF}  +2r_H \frac{\partial a}{\partial \ell}\right.\nonumber\\
& \left. - 2\frac{\partial}{\partial \ell}\ln| e^{r_H (a-b)}- 1|  - 2\frac{\partial}{\partial \ell}  \ln | e^{r_H(a+b)}- 1| \RT.
\end{align}
After integration over $\zeta$, the first two terms vanish since they are total derivatives and $F'$ is periodic. Moreover, note that $F(0)=0$, $F(2\pi)=2\pi$ and $F'$ is periodic. This means that $F$ can be written as $F(\vp)=\vp+\Fcal(\vp)$ where $\Fcal$ is periodic. Using this fact and the expression for $a$ given by \eqref{eq:EJeq1} it is easy to show that the third term vanishes as well upon integration over $\zeta$.
Thus the differential entropy becomes
\begin{align}
4G_N\Ecal = \frac{\partial}{\partial \ell} \int\limits_0^{2\pi} d\zeta \ln | e^{r_H (a-b)}- 1| + \frac{\partial}{\partial \ell} \int\limits_0^{2\pi} d\zeta \ln | e^{r_H (a+b)}- 1|.\label{eq:Ecal_VBTZ_1}
\end{align}
From \eqref{eq:EJeq1} and \eqref{eq:EJeq2}, we have $a-b=F(-t+\zeta+\ell)-F(-t+\zeta)$ and $a+b=F(t+\zeta)-F(t+\zeta+\ell)$. These are both periodic functions of $\zeta$ since the linear dependence on $\zeta$ in $F$ cancels. Moreover, $a-b$ is a function of $\zeta-t$ and $a+b$ is a function of $\zeta+t$. Thus, since they are both periodic in $\zeta$, we can shift the integration variable $\zeta$ by $\zeta\rightarrow\zeta+t$ or $\zeta\rightarrow\zeta-t$ in the integrals in \eqref{eq:Ecal_VBTZ_1} to get rid of the time dependence. Note also that $a-b>0$ and $a+b<0$ due to the monotonicity of $F$. Putting all this together we obtain
\begin{align}
4G_N\Ecal &= \frac{\partial}{\partial \ell} \int\limits_0^{2\pi} d\zeta \ln \LF e^{r_H (F(\zeta+\ell)-F(\zeta))}- 1\RF + \frac{\partial}{\partial \ell} \int\limits_0^{2\pi} d\zeta \ln \LF 1-e^{r_H (F(\zeta)-F(\zeta+\ell))}\RF\nonumber\\
&=2\frac{\partial}{\partial \ell} \int\limits_0^{2\pi} d\zeta \ln \sinh\LF\frac{r_H(F(\zeta+\ell)-F(\zeta))}{2}\RF,\label{eq:Ecal_VBTZ}
\end{align}
which in particular shows that the differential entropy is time-independent, so in some sense the integration over $\zeta$ averages out the variations in $\partial_\ell S$. 
\subsection{Inhomogeneous shell}\label{sec:shell_non_pert}
In this section we explain how to numerically compute entanglement entropy and differential entropy in the presence of an inhomogeneous shell of arbitrary energy density.

Geodesics in the shell background satisfy the \ads equations of motion when $v<0$ and the \VBTZ ones when $v>0$. As we have emphasized many times in this article, there exists a coordinate transform from $(v,r,\phim)$ to $(v,\rp,\phip)$ that locally maps \VBTZ into a static BTZ and accordingly the geodesic solutions at $v>0$ are locally those of static BTZ. Each geodesic consists of a connected \ads piece with two endpoints on the shell and two pieces in the BTZ part each connecting a point on the shell with a point on the asymptotic boundary. The endpoints on the boundary in static BTZ coordinates are again determined in terms of the $(\vp,t)$ boundary coordinates by \eqref{eq:boundary_points}.
To fix the geodesic completely one needs to know how the three pieces are connected to each other. First of all, we demand that the geodesic is continuous across the shell. The coordinate derivatives are then fixed by refraction conditions which follow from extremizing the geodesic action with respect to the shell crossing points. The resulting refraction conditions are
\begin{align}
\dot{v}_{\text{\VBTZ}} &= \dot{v}_{\text{\ads}}F'\LF \vp\RF,\label{refr_1}\\
r^{2}\dot{\vp}_{\text{\ads}} &= r^{2}\dot{\vp}_{\text{\VBTZ}} - r\dot{v}_{\text{\VBTZ}} \frac{F^{''}\LF \vp\RF}{F^{'2}\LF \vp\RF}\label{refr_2}.
\end{align}
and are derived in appendix~\ref{app:inhom_refrac}. For ease of notation we have suppressed indices, but one should keep in mind that these conditions are only valid at the two shell crossing points. These conditions ensure that the geodesic is extremal in the whole spacetime and not just in the two patches separately.\\

We now want to construct a geodesic which is anchored at the boundary points $(\phip^{(i)},\tp^{(i)})$. The two pieces of geodesic in the BTZ part ($v>0$) have free parameters $E_1,J_1,\vp_{0,1},v_{0,1}$ and $E_2,J_2,\vp_{0,2},v_{0,2}$ respectively (see section \ref{sec:btzgeo}), while that in the AdS part ($v<0$) has free parameters $\eads,\jads,\vp_{0,\text{AdS}},v_{0,\text{AdS}}$ (see section \ref{sec:adsgeo}). To construct such a geodesic, we start with a guess for $E_1$ and $J_1$, while $\vp_{0,1}$ and $v_{0,1}$ are fixed by demanding the geodesic to be anchored at $(\phip^{(1)},\tp^{(1)})$. By using continuity and the refraction conditions \eqref{refr_1}-\eqref{refr_2} at the crossing point where this geodesic intersects the shell at $v=0$, we can determine all free parameters $\eads,\jads,\vp_{0,\text{AdS}},v_{0,\text{AdS}}$ for the AdS piece. By then enforcing continuity and the refraction conditions \eqref{refr_1}-\eqref{refr_2} again at the second crossing point, we can determine all free parameters $E_2,J_2,\vp_{0,2},v_{0,2}$ for the second BTZ piece. We now want the second BTZ geodesic to reach the boundary at the point $(\phip^{(2)},\tp^{(2)})$, and we need to choose $E_1$ and $J_1$ appropriately such that this is the case. Since it seems difficult to solve this analytically for $E_1$ and $J_1$, we employ a numerical root finding algorithm to tune $E_1$ and $J_1$ such that the second geodesic reaches the boundary at the point $(\phip^{(2)},\tp^{(2)})$. This is done at every time $t$ we are interested in and as initial guess to this root finding algorithm we can use the $E_1$ and $J_1$ found at a previous time $t-\Delta t$. The initial guess at times close to zero can be obtained by approximating $\eads$ and $\jads$ by their values for $v<0$ (which can be found analytically, see section \ref{sec:adsgeo}), and then using the refraction conditions to obtain an initial guess for $E_1$ and $J_1$. This will be used in the beginning of the procedure (the first time we run the root finding algorithm) where no previous values for $E_1$ and $J_1$ are available.

\subsubsection{Entanglement entropy}
Once the geodesic solution is known, its length is easily found by summing the length of the two legs in BTZ and the leg in \ads. By the HRT prescription, the holographic entanglement entropy follows as the regularized length of the geodesic. We will discuss its behaviour over time in more detail in section~\ref{sec:examples} for specific profiles of the stress-energy tensor. 
\subsubsection{Differential entropy}
Entanglement entropy is only known numerically for a given opening size $\ell$. To compute differential entropy from the definition~(\ref{eq:Ecal}) one needs to know the derivative of $S$ with respect to $\ell$. This can be done numerically, but is time consuming. Instead we apply the equivalent definition~\eqref{eq:Ecal_Myers}. In~\cite{Headrick:2014eia} it was shown that~(\ref{eq:Ecal_Myers}) is constant along the geodesic, so we choose to evaluate it at the endpoint $\la=\la_2$. In BTZ coordinates then~\eqref{eq:Ecal_Myers} reduces to
\begin{equation}
\Ecal\LF \ell,t\RF = \oint d\zeta\LT \frac{dt^{(2)}}{d\ell} E_2 + \frac{d\phip^{(2)}}{d\ell} J_2 + \frac{dr^{(2)}_{c}}{d\ell} \dot{\rp}(\la_2)\RT .
\end{equation}
For any given $\ell$ the energy and angular momentum can be determined and the asymptotic boundary transform allows to compute the derivatives of $t^{(2)}$ and $\phip^{(2)}$ with respect to $\ell$. Furthermore by using \eqref{eq:r_cutoff} and the BTZ equation of motion it can be shown that the term proportional to $\dot{\rp}$ vanishes after integration over $\zeta$.  This way $\Ecal$ can be computed in a more efficient way than by numerically differentiating $S$ with respect to $\ell$. The integration over $\zeta$ still needs to be performed numerically\footnote{We use a trapezoidal routine to perform the integration, which means that we determine a vector of values for the integrand and then compute the integral as a sum involving the elements in this vector.}.

\section{Examples}\label{sec:examples}
Given any holographic CFT stress-energy tensor satisfying $T_\pm(y_\pm )= T(y_\pm)$, we can use the machinery explained in this paper to compute the entanglement entropy for any interval on the boundary. The behaviour of the entanglement entropy depends quite a lot on the profile of the stress-energy tensor, hence we restrict ourselves to discussing its behaviour in two particular examples. One is an oscillatory stress-energy tensor that could be interesting for quenches in for example trapped cold atom systems, but the main motivation for this example is to have some generic inhomogeneity on top of a homogeneous quench so that in particular we can compare the perturbative results to the non-perturbative results. The other is a bilocal quench formed by two delta function peaks in the stress-energy tensor, and we will also consider a smooth version of this setup where the delta functions have a finite width. We will use these examples to extract some generic features of entanglement entropy in inhomogeneous quenches and to point out interesting behaviour that can show up, but we will not perform an in-depth study to extract all possible features.
\subsection{Oscillatory quench}

\begin{figure}
\centering
\includegraphics[scale=0.5]{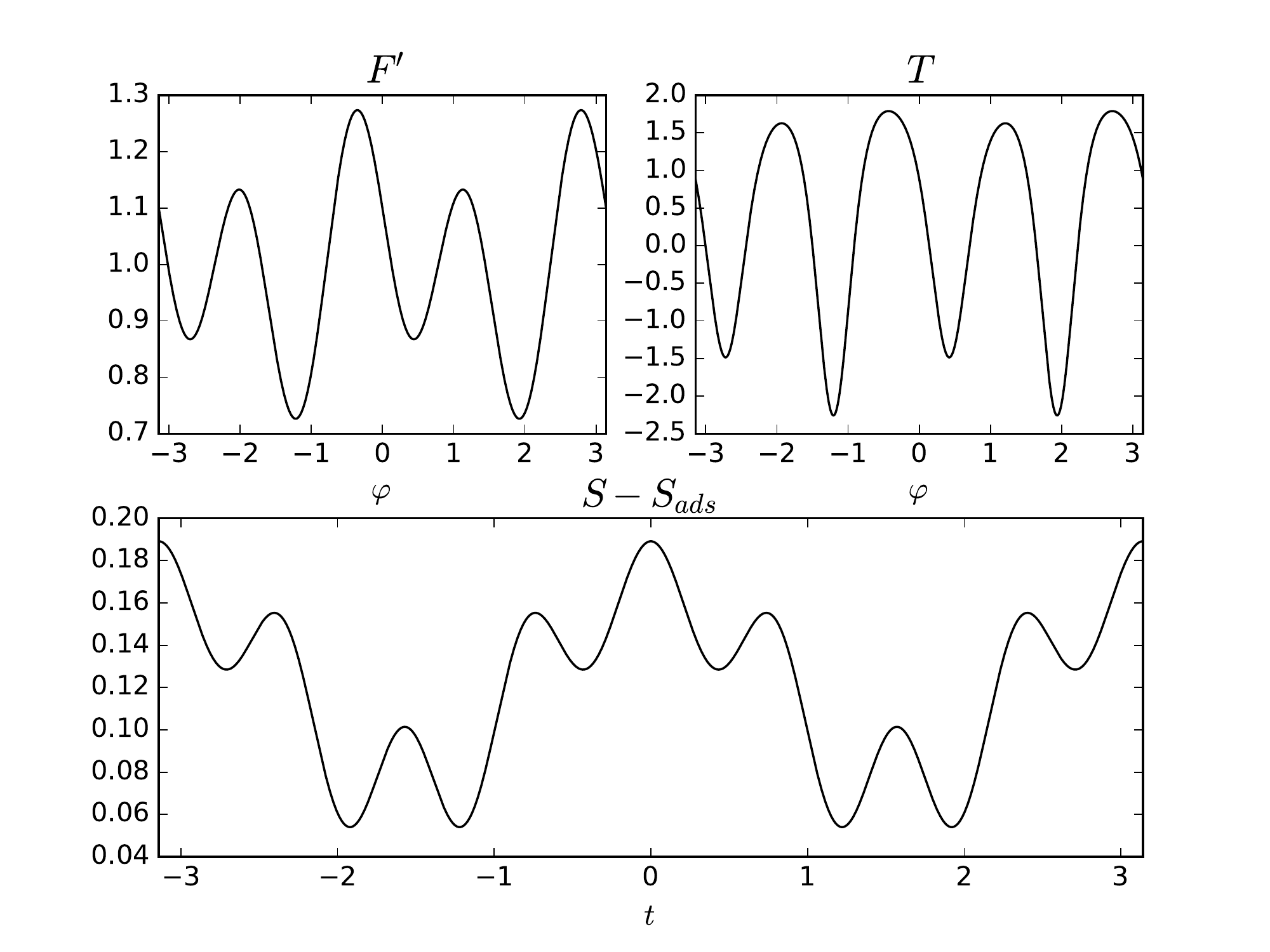} 
\caption{\label{fig_inhom_entropy} The function $F'$ as given by \eqref{F_inhom} with $\epsilon=0.05$, the corresponding stress-energy tensor $T$ and the entanglement entropy of an arbitrary interval located at $(-0.4,0.4)$ in the eternal \VBTZ spacetime. The plotted range is $(-\pi,\pi)$ but all these three plots should be continued periodically.}
\end{figure}

In this section we will consider a quench where the inhomogeneous part of the function $F$ is composed of some trigonometric functions with the aim of studying properties of a quench with some generic inhomogeneities added on top of a homogeneous background. The particular profile for $F$ we will choose is
\begin{equation}
F(\vp)=\vp+\e (\sin(2\vp)+\cos(4\vp)),\label{F_inhom}
\end{equation}
but the exact profile in this case is not very important. The parameter $\epsilon$ determines the strength of the inhomogeneity and we can compare small $\e$ results with our perturbative results. The corresponding boundary stress-energy tensor $T$ is then given by \eqref{stress_energy_CFT_eq}, and is shown in Figure \ref{fig_inhom_entropy}. For $v<0$ the stress-energy tensor is that of \ads, namely $T=-1/4$. 

\begin{figure}
\centering
\includegraphics[scale=0.6]{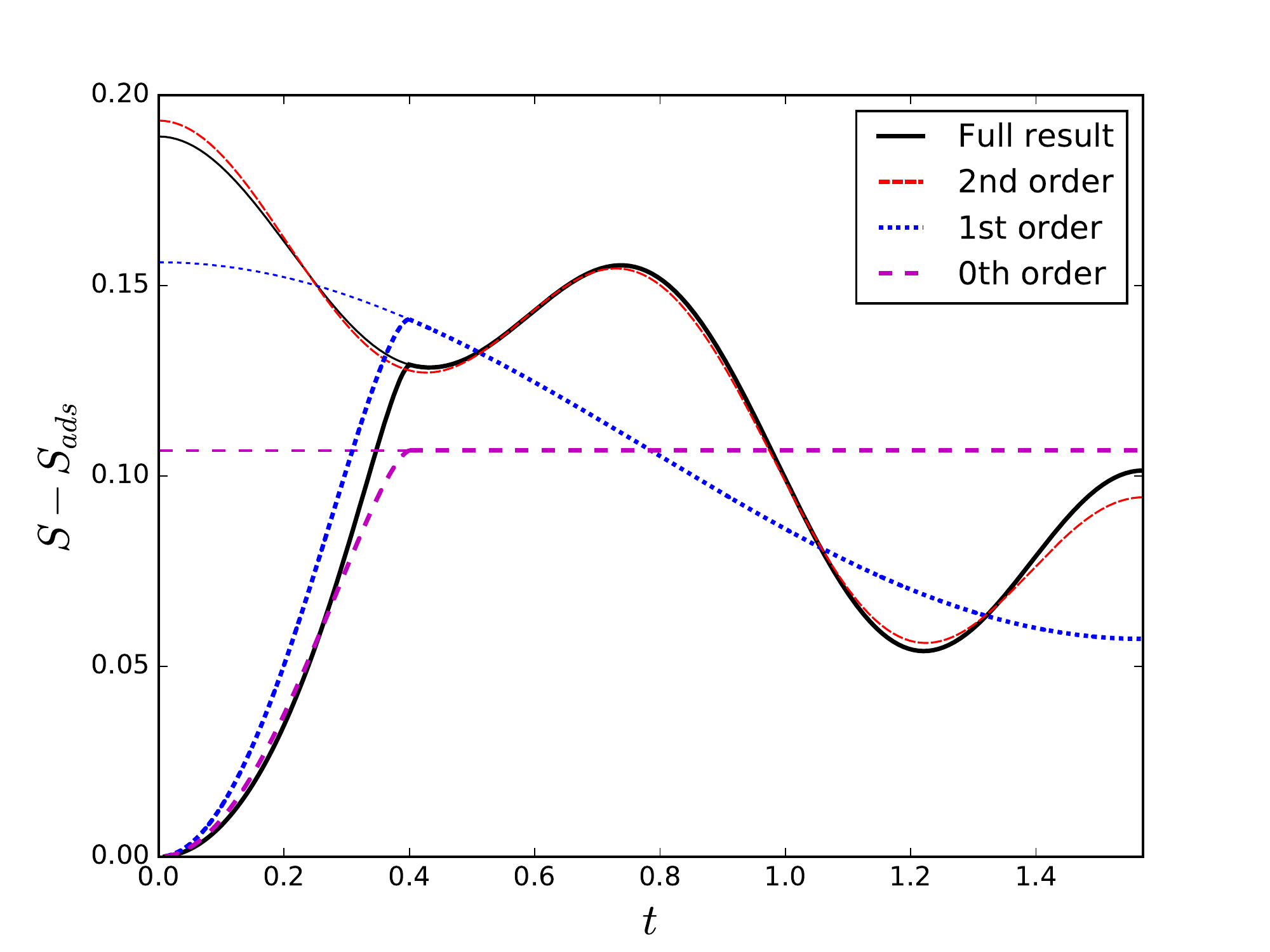} 
\caption{\label{fig_inhom_entropy_2} The dynamical evolution of the entanglement entropy of the interval $(-0.4,0.4)$ after a global quench with inhomogeneity given by \eqref{F_inhom} with $\epsilon=0.05$, comparing the full numerical result with perturbation theory. The entanglement entropy smoothly interpolates from the value in \ads to the late time \VBTZ result. Entanglement entropy in the eternal \VBTZ solution is shown with thinner lines while entanglement entropy with the shell is shown with thicker lines. The second order result is only showed for eternal \VBTZ\,\hspace{-2mm}.}
\end{figure}

\begin{figure}
\centering
\includegraphics[scale=0.6]{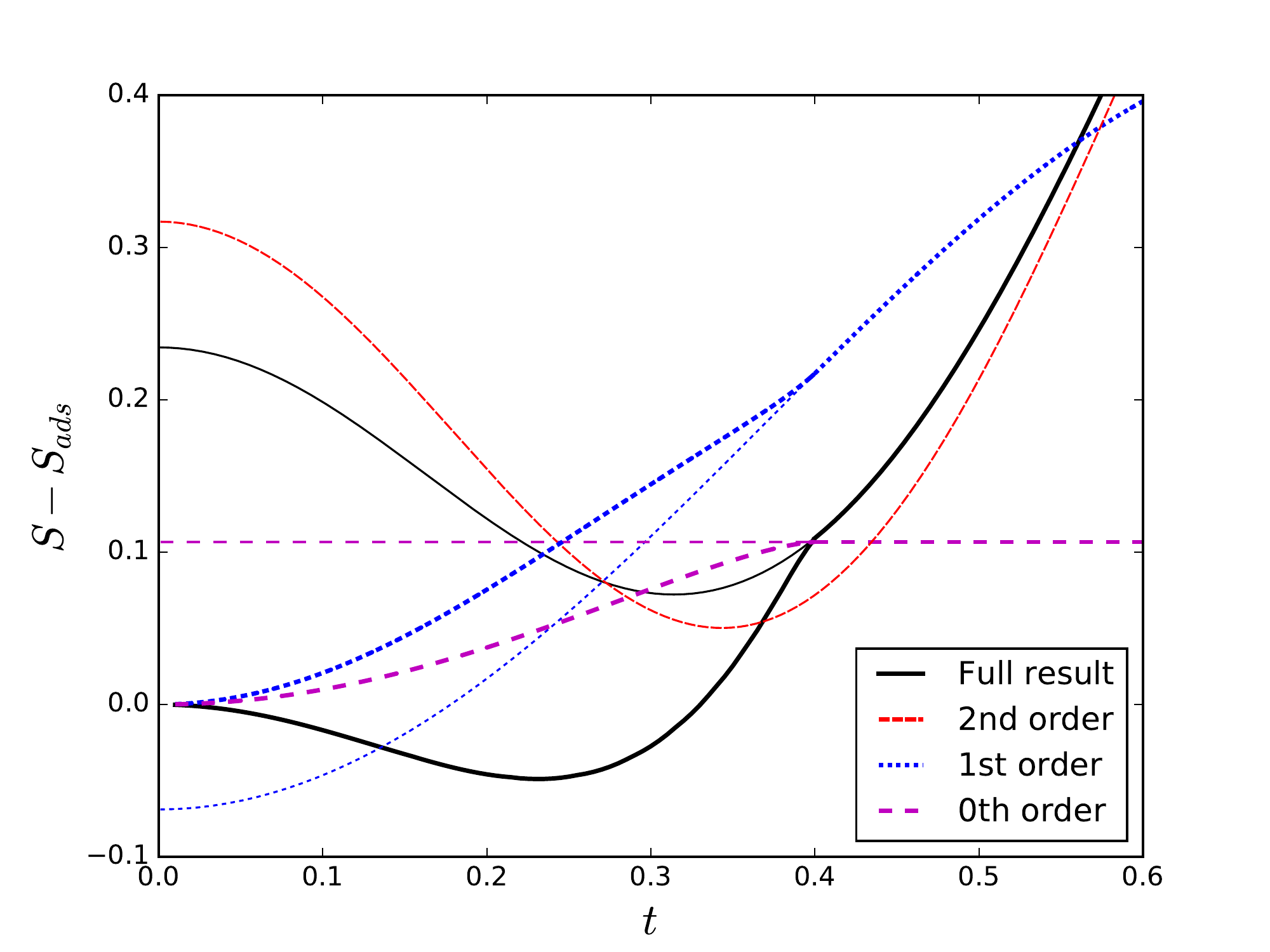} 
\caption{\label{fig_inhom_entropy_2b} The dynamical evolution of the entanglement entropy of the interval $(-0.3,0.5)$ after a global quench with inhomogeneity given by \eqref{F_inhom} with $\epsilon=0.15$, comparing the full numerical result with perturbation theory. The entanglement entropy smoothly interpolates from the value in \ads to the late time \VBTZ result. Entanglement entropy in the eternal \VBTZ solution is shown with thinner lines while entanglement entropy with the shell is shown with thicker lines. The second order result is only showed for eternal \VBTZ\,\hspace{-2mm}.}
\end{figure}

The entanglement entropy, compared with the perturbative result, is given in figure \ref{fig_inhom_entropy_2} and \ref{fig_inhom_entropy_2b} for different values of $\e$. Generically, the entanglement entropy starts out at its value in \ads at $t=0$ and smoothly transitions to the late time solution. The saturation time $t_{\text{sat}}$, which is the time until the geodesic no longer intersects the shell, is equal to $\ell/2$ as we show in appendix \ref{app:saturation}. This time scale is unrelated to any time scales associated with the time dependence of the stress-energy tensor. After saturation at $t=\ell/2$, the entanglement entropy equals that of $\text{\VBTZ}$. The final entropy, as illustrated in figure \ref{fig_inhom_entropy}, is clearly not constant but oscillates over time. The absence of dissipation in $d=1+1$ CFTs prohibits the entanglement entropy from thermalizing. Instead we should interpret the result as the entropy of a steady state. Although it is not clear at this stage how it comes about from the microscopics, we view the smooth transition from vacuum to \VBTZ as a prethermalization. By prethermalization we loosely mean here the ``fast" relaxation towards a steady state, in line with other definitions \cite{Berges:2004ce,Langen:2016vdb}. Note that in figure \ref{fig_inhom_entropy_2b}, the entropy first decreases from its \ads value. Whether the entropy increases or decreases during the prethermalization phase depends on the strength and shape of the inhomogeneity as well as the location and size of the interval. Figures \ref{fig_inhom_entropy_2} and \ref{fig_inhom_entropy_2b} also show the entanglement entropy perturbatively in $\e$ around the homogeneous shell background compared to the exact numerical result. To zeroth order the entropy grows until it saturates at $t=\ell/2$ and is then constant. The first order corrections do not coincide well with the numerical result, because at late times they only get contributions from the odd part of $F$. In figure \ref{fig_inhom_entropy_2}, we see that to second order in $\e$ the late time entropy agrees quite well with the numerical result. In figure \ref{fig_inhom_entropy_2b}, where $\e$ is larger, the second order correction is not very accurate either.\\

\subsubsection{Differential entropy}
\begin{figure}[h]
\centering
\includegraphics[scale=0.4]{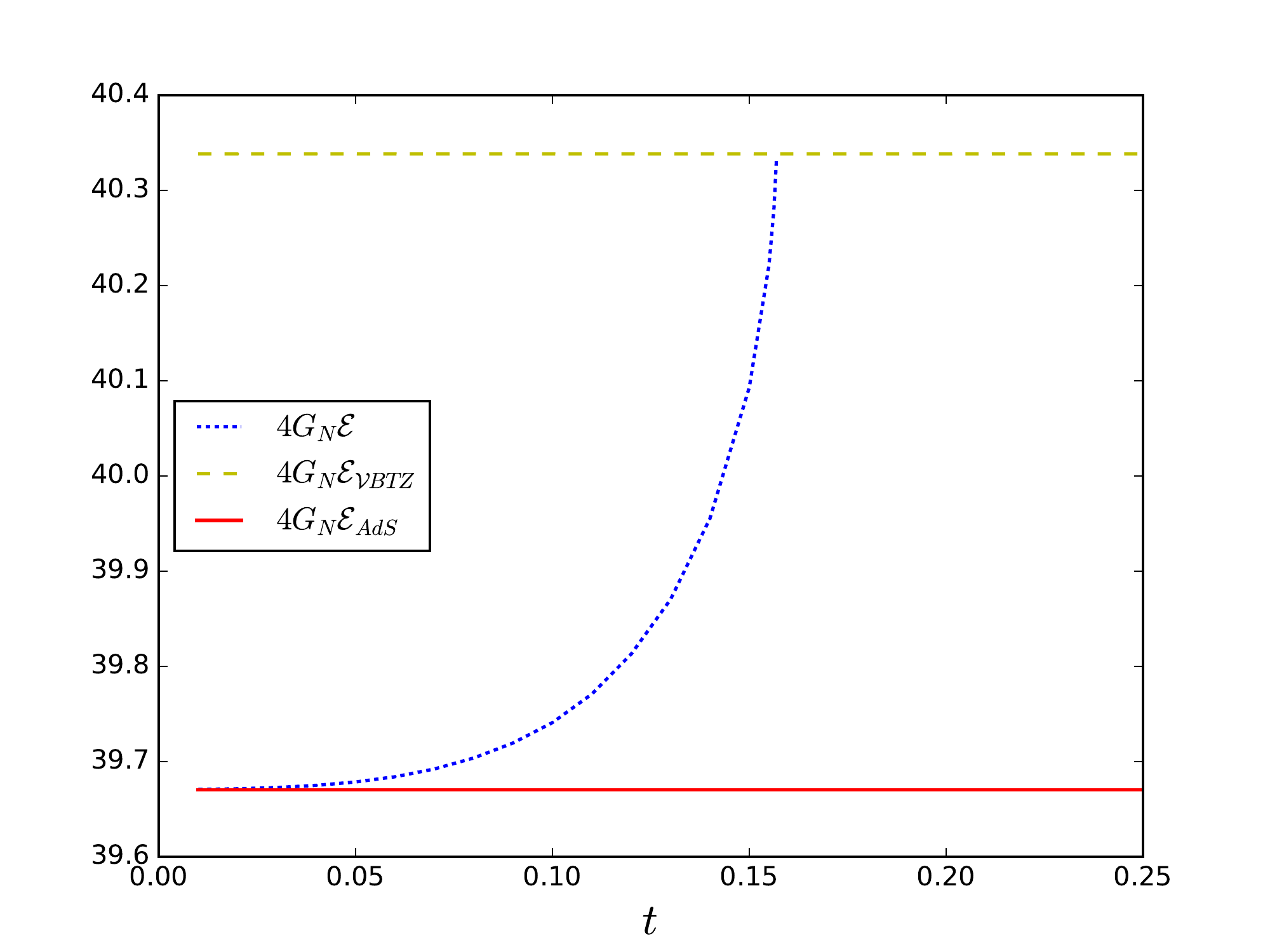}
\caption{\label{fig:diff_ent_inhom} $\Ecal(t)$ of the oscillatory quench background (blue, short dashes). We have also plotted the vacuum differential entropy (red, solid) and in \VBTZ $ $ (yellow, long dashes). The vacuum differential entropy is plotted for times $0<t<\ell/2$, while the $\Ecal_{\text{\VBTZ}}$ is plotted until $t=\ell/2+0.1$. The interval size here is $\ell=\pi/10$.}
\end{figure}

The differential entropy $\Ecal (t)$ behaves analogously to the homogeneous shell background. It starts out at its vacuum value and smoothly grows until it saturates at $t=\ell/2$ at its value in \VBTZ . From then onwards the differential entropy is constant, as we have analytically shown in section \ref{sec:final_non_pert}. Figure \ref{fig:diff_ent_inhom} plots $\Ecal (t)$ for $\ell = \pi /10$. It saturates at $t = \ell/2 \approx 0.15$ and then evolves according to the \VBTZ $ $ differential entropy which is constant.
For larger intervals, the behaviour of differential entropy is qualitatively similar but the vacuum value is pushed downwards towards $\Ecal (t)=0$. In the limit of $\ell=\pi$ it becomes hard to numerically solve for the geodesics, so we cannot study the formation of the entanglement shadow. Nevertheless we suspect an entanglement shadow will form because of the similarity with the homogeneous quench for all intervals that we can plot. Secondly, for perturbative inhomogeneities the entanglement shadow is that of homogeneous Vaidya, at least to first order in $\e$. 
\subsection{Bilocal quench}\label{bilocal}
In this section we will study a bilocal quench, which consists of two delta function quenches in the boundary CFT, and the bulk dual will be that of two point particles that are created at the boundary which then fall into the bulk\footnote{Entanglement entropies in the bilocal quench geometry have already been studied in \cite{Arefeva:2017pho}, however in their work the authors choose coordinates under the assumption that the final geometry is a static BTZ black hole. As we have emphasized, the final state is uniquely given by the time dependent \VBTZ geometry, which explains the discrepancy between our and their results. We thus believe that in the coordinates chosen in \cite{Arefeva:2017pho}, the spacetime before the shell would not be the standard anti-de Sitter space, but rather a dressed \ads (\ads on which one has applied a large gauge transformation, and thus has non-constant boundary stress-energy tensor modes) which is fine tuned such that the late time geometry exactly matches that of a static BTZ black hole.}. This spacetime was first constructed in \cite{Matschull:1998rv} where the particles collide at the center to form a black hole solution\footnote{Note that the fact that the particles collide, instead of miss each other, can be seen as a very fine tuned setup, but since this happens behind the event horizon anyway we expect that boundary observables will not change if the particles instead miss each other.}. This can be formulated in our general framework by just choosing an appropriate function $F(\vp)$ for which the stress-energy tensor has two delta function singularities. For times $v>0$, the stress-energy tensor consists of right- and left-moving pointlike excitations emanating from the location of the bilocal quench. A derivation of the function $F$ from the original setup in \cite{Matschull:1998rv}, as well as a lightning review of that setup, is given in Appendix \ref{2p_appendix}. The transformation $F$ can be found by integrating
\begin{equation}
F'(\vp)= \left\{
\begin{array}{cc}
\frac{1}{\sqrt{M}} \frac{\sinh\LF \frac{\pi \sqrt{M}}{2}\RF}{-\cosh\LF \frac{\pi \sqrt{M}}{2}\RF\sin\LF \vp \RF +1},& -\pi/2\leq\vp\leq\pi/2, \\
                       \frac{1}{\sqrt{M}} \frac{\sinh\LF \frac{\pi \sqrt{M}}{2}\RF}{\cosh\LF \frac{\pi \sqrt{M}}{2}\RF\sin\LF \vp \RF +1},&\pi/2\leq\vp\leq3\pi/2,\\
\end{array}\right.
\end{equation}
extended periodically to all other angles. The local quenches are located at $\vp=\pm\pi/2$, which will be the case throughout this section. We will also choose the energies of the particles such that the mass of the final black hole is $M=1$.

\begin{figure}[h]
\centering
\includegraphics[scale=0.5]{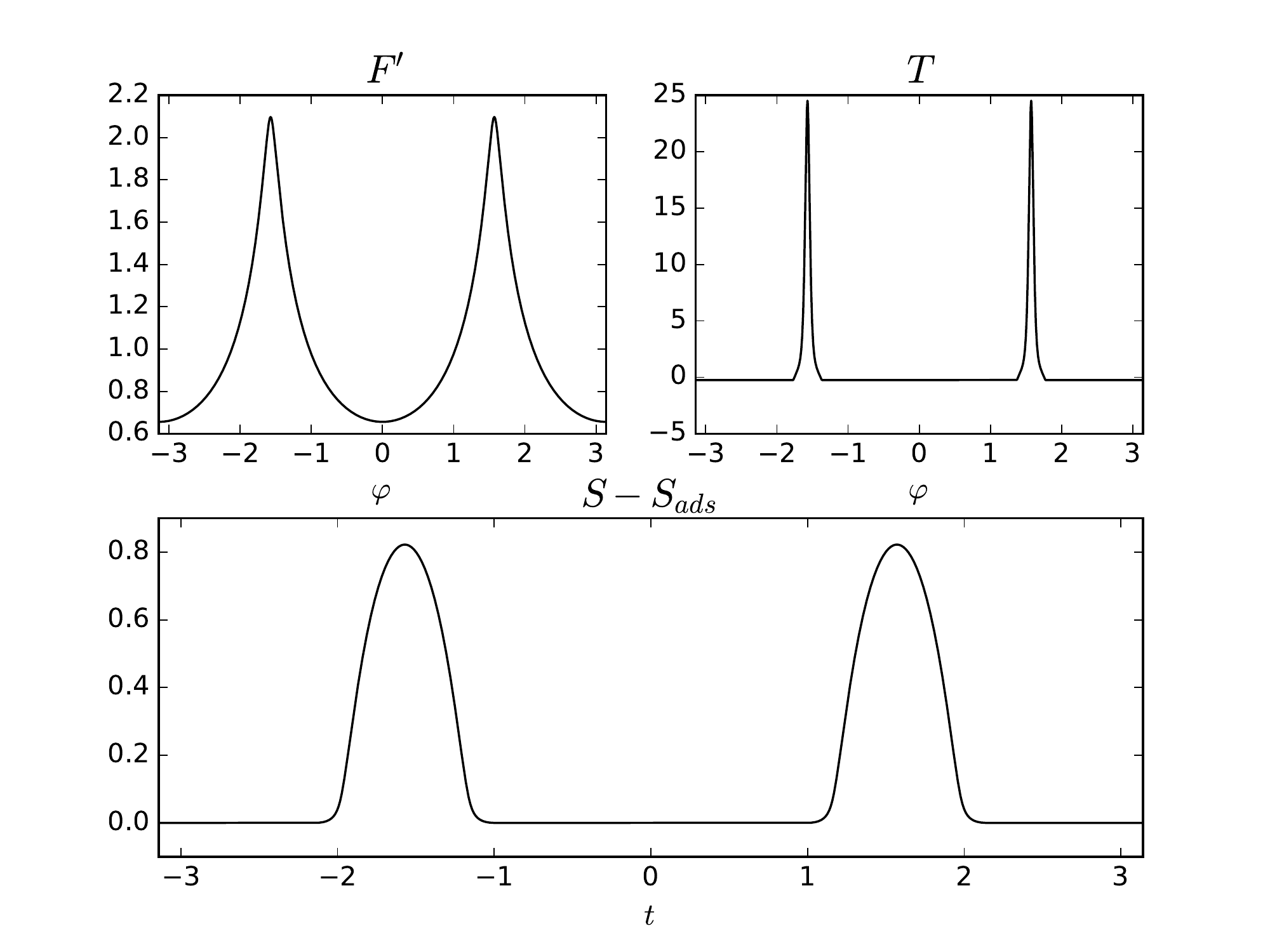} 
\caption{\label{fig_2particles_5}
The function $F'$, the corresponding stress-energy tensor $T$ and the entanglement entropy of an arbitrary interval located at $(-0.4,0.4)$ in the eternal \VBTZ spacetime, where $F'$ is the one that would result from the two-particle collision process. The plotted range is $(-\pi,\pi)$ but all these three plots should be continued periodically. Note that $F'$ has been smoothened out according to equation \eqref{smooth_kink} with $\sigma=0.04$.}
\end{figure}

\begin{figure}[h]
\centering
\includegraphics[scale=0.5]{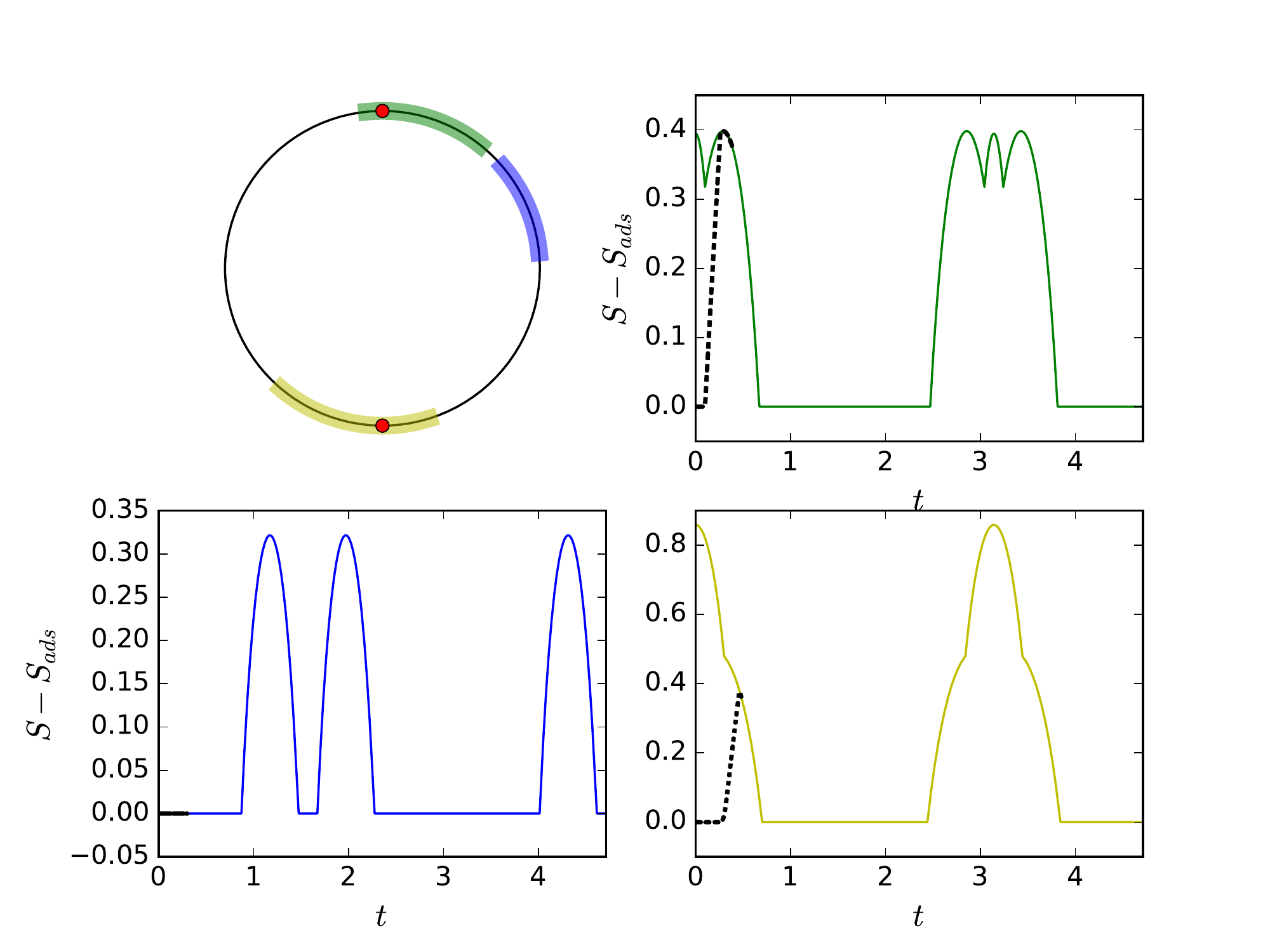} 
\caption{\label{fig_2particles_1}Entanglement entropy in the late time geometry for three different intervals in the bilocal quench. The energies of the particles (indicated by the red dots) are all such that the final black hole mass is $M=1$. The topmost interval (green), which is the interval $(0.9,1.67)$, corresponds to the top-right plot, the right interval (blue), which is the interval $(0.1,0.7)$, corresponds to the down-left plot and the downmost interval (yellow), which is the interval $(-\pi/2-0.7,-\pi/2+0.3)$, corresponds to the down-right plot.  The colored solid lines correspond to the eternal \VBTZ result, while the transition from \ads to the \VBTZ result in the presence of the shell is shown with black dashed lines, from $t=0$ until the time $\ell/2$ where the geodesic no longer intersects the shell. For other times all the colored solid lines should be continued periodically.}
\end{figure}

\begin{figure}[h]
\centering
\includegraphics[scale=0.5]{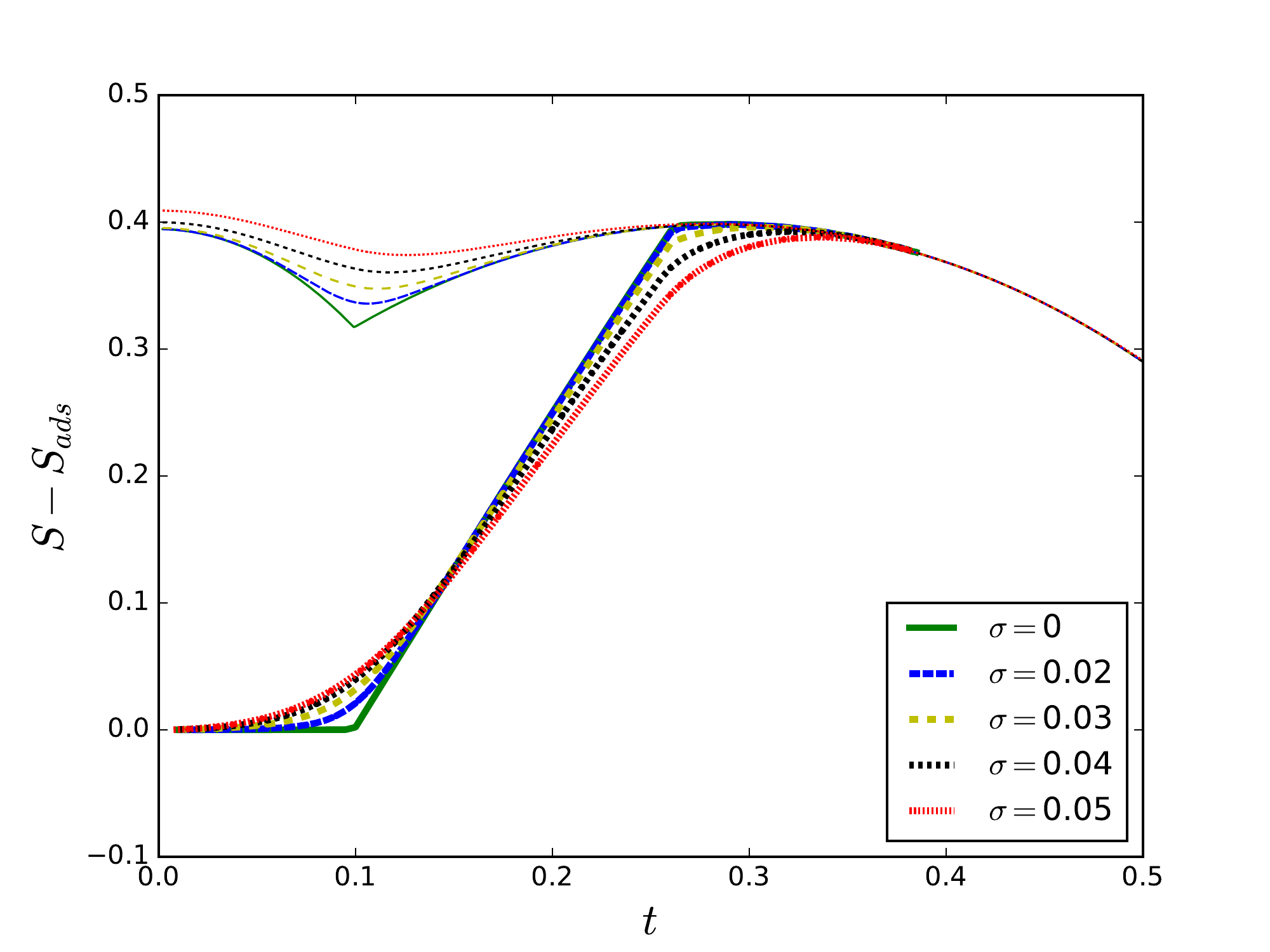} 
\caption{\label{fig_2particles_2}Entanglement entropy for the bilocal quench for the interval $(0.9,1.67)$, at different widths $\sigma$ as defined by equation \eqref{smooth_kink}. The entanglement entropy starts out at the value in \ads and interpolates to the value in ethernal \VBTZ at $t=\ell/2$ and equals that of eternal \VBTZ for $t>\ell/2$. The entanglement entropy for eternal \VBTZ is also shown for all times. Smoother curves correspond to larger $\sigma$.}
\end{figure}

\begin{figure}[h]
\centering
\includegraphics[scale=0.5]{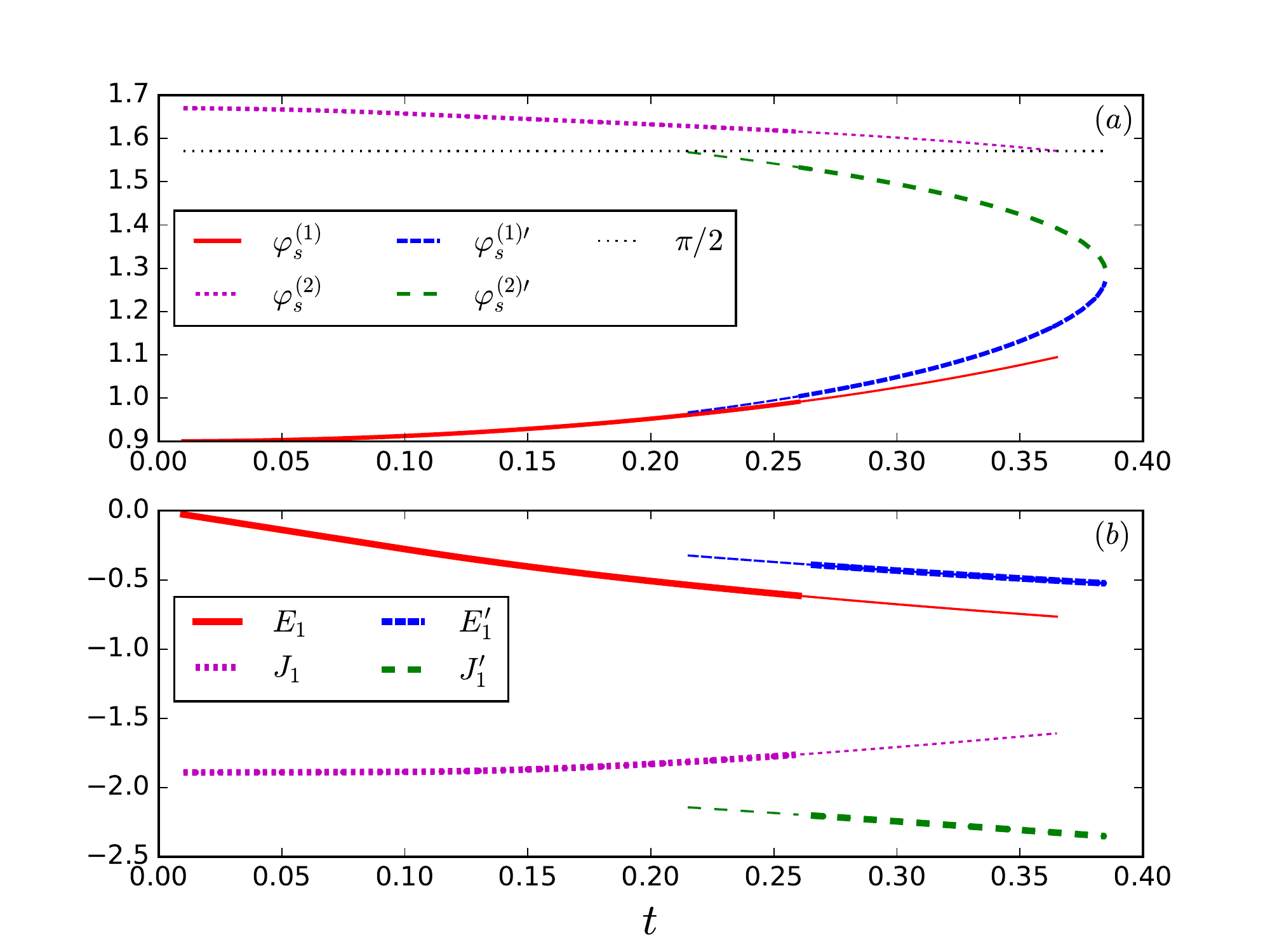} 
\caption{ \label{fig_2particles_4} $(a)$: Angular location of the crossing points of the two competing geodesics in the bilocal quench geometry, where one of the competing geodesics is marked with a prime. $(b)$: $E_1$ and $J_1$ for the two competing geodesics in the bilocal quench. The thick lines correspond to the geodesic that at that time has the smallest length. The interval used here is $(0.9,1.67)$.}
\end{figure}

In figure \ref{fig_2particles_1} we illustrate how entanglement entropy evolves through time for three different intervals. The qualitative behaviour depends on the location and size of the interval. Let us first comment on the final state. Notice that the entropy of \VBTZ contains plateau phases at which it equals the \ads entropy. To explain this, recall that after the quench, the stress-energy tensor modes will consist of two left-moving and two right-moving delta function excitations, emerging at the speed of light from the locations of the two local quenches. Roughly speaking, we see that the entanglement entropy of an interval is larger if there are more of these excitations located inside the interval. If there are no excitations inside the interval, the entropy is insensitive to the quench and equal to that of the vacuum. Whenever a delta function excitation crosses the interval, the entropy changes (continuously, albeit not smoothly).

In the presence of a bilocal quench at $v=0$, entanglement entropy generically starts out at its vacuum value and interpolates monotonously to its value in the \VBTZ spacetime which it acquires at some time $t\leq\ell/2$. The phase where the geodesic crosses the shell, where the entanglement entropy interpolates from the value in \ads to the value in \VBTZ, is illustrated with black lines in figure \ref{fig_2particles_1}. During a short initial time phase, the entanglement entropy is equal to that of \ads until at least one of the stress-energy tensor excitations has crossed one of the endpoints of the interval. In particular, when none of the quench locations are inside the interval (as in the lower left plot of figure \ref{fig_2particles_1}) the entropy immediately connects the value in vacuum \ads to that of \VBTZ in a trivial way because both values are equal and the quench is locally trivial. In the two right plots of figure \ref{fig_2particles_1} we see a growth of entropy which interpolates between the \ads value and the \VBTZ value in two cases where one of the quench locations is inside the interval. Note that even though the time when the geodesic leaves the shell will always be at $t=\ell/2$ (as we prove in appendix \ref{app:saturation}), the entropy can saturate to that of \VBTZ at some time $t\leq \ell/2$. This is because the shell in this example only contains energy density at $\vp=\pm\pi/2$ and is empty everywhere else, and the statement that the geodesic crosses the $v=0$ surface does not need to have any physical meaning.\\

It is also possible to consider a smoother version of $F$, where we replace the kinks in $F'$ at $\pm \pi/2$ by a smoother function, and we will use the function
\begin{equation}
\Theta(\vp)=A\log(\cosh(\vp/\sigma))+B+C\vp^2.\label{smooth_kink}
\end{equation}
To be more specific, given some small parameter $\delta$, we replace $F'$ in the intervals $(\pm\pi/2-\delta,\pm\pi/2+\delta)$ by $\Theta(\vp\mp\pi/2)$. The parameters $A$, $B$ and $C$ are then fixed by matching the values, first and second derivatives at $\vp=\pm\pi/2+\delta$ and $\vp=\pm\pi/2-\delta$. The parameter $\sigma$ sets the width (or smoothness) of the kink. For convenience we fix $\delta=5\sigma$. The illustrations in figure \ref{fig_2particles_5} of the stress-energy tensor function $T$, $F'$ and entanglement entropy for the final state have $\sigma=0.04$. In Figure \ref{fig_2particles_2} we show the full entanglement entropy for several values of $\sigma$ where we clearly see that the entanglement entropy becomes smoother for larger $\sigma$. \\

In certain cases there will exist two geodesics that exchange dominance. The reason for this is that the infalling particle, which is essentially a lightlike conical deficit, modifies the spacetime such that two geodesics connected to the same boundary points can exist at the same time, where one geodesic goes around the particle and one does not. This transition is illustrated in figure \ref{fig_2particles_4} for the interval $(0.9,1.67)$ and for $\sigma=0$, where we plot the values of $E_1$ and $J_1$ (energy and angular momentum for the first BTZ part of the geodesic) as well as the angles where the geodesic crosses the shell at $v=0$. The geodesic that does not go around the particle is denoted with a prime, and the currently dominating geodesic (the one with shorter length) has been marked in bold. The second (primed) geodesic only exists after $t\approx 0.215$, when the second crossing angle $\vp_{s}'^{(2)}$ of the geodesic equals to $\pi/2$ (which is the location of one of the infalling point particles). Similarly, the first (unprimed) geodesic disappears exactly when its second crossing angle $\vp_{s}^{(2)}$ hits $\pi/2$ at $t\approx0.365$. At $t\approx 0.26$ the second (primed) geodesic becomes the dominant (shorter) one and should thus be used when computing the entanglement entropy. The exchange of dominance of these two geodesics is what causes the non-smooth evolution of the entanglement entropy at $t\approx 0.29$ for $\sigma=0$ in figure \ref{fig_2particles_2}. Furthermore, the second geodesic is essentially the same geodesic as the one we would have in eternal \VBTZ and in particular the values of $E_1$ and $J_1$ are the same. The fact that it still crosses the shell until $t=\ell/2$ has no physical significance since the shell is empty there (no energy density).

\section{Summary and conclusions}\label{sec:conclusions}
In this paper we have examined the entanglement entropy and differential entropy in inhomogeneous quenches in $d=1+1$ holographic CFTs modeled by a massless shell falling in from the boundary in the dual spacetime. The shell can be viewed as an infinite number of collapsing point particles and by tuning the energy density, we can study arbitrary spinless stress-energy tensors. The dual geometry is that of \ads on one side of the shell, glued to an inhomogeneous black hole spacetime, \VBTZ\,\hspace{-2mm}, which can be constructed by applying a large gauge transformation to the BTZ geometry. We have constructed geodesics in these spacetimes. By using the fact that \VBTZ can be mapped to a static BTZ geometry by a coordinate transformation, these computations essentially reduce to gluing together geodesics in \ads and in BTZ via suitable refraction conditions when crossing the shell.\\

Perturbatively in the amplitude of the inhomogeneities, the holographic entanglement entropy can be computed analytically. In the eternal \VBTZ geometry, we can even treat the entropy analytically for any non-perturbative inhomogeneity. At first order, the entropy is a linear combination of left- and right moving modes. At non-perturbative inhomogeneity in \VBTZ it can be expressed in terms of a simple function of left- and right movers as well. Because of the  absence of dissipation in $2+1$-dimensional gravity, the entropy generically shows non-trivial dynamical behaviour at arbitrary large times. The \VBTZ geometry therefore does not really describe a thermal state but rather a steady state. \\

The perturbative entanglement entropy in the final state agrees with a CFT quench computation of \cite{CardySotiriadis2008}. These authors have also shown that when the scale of spatial variation is small, the entropy can be reproduced from a quasi particle picture. We can also show that the perturbative final state entropy can be reproduced as a variation in the \VBTZ frame leading to a formula that resembles a first law of entanglement entropy.\\

In the presence of the inhomogeneous shell, we numerically solve for the geodesics, whose lengths compute the entanglement entropy of a boundary interval. We distinguish three regimes. At $t<0$ the entropy is that of the vacuum and the geodesics are just that of \ads. Once the shell is injected from the boundary at $t=0$, the geodesics will contain one piece in the \ads part of the geometry at $v<0$, and two pieces in the \VBTZ part of the geometry. They cross the shell as long as $t<\ell/2$ and saturate at $t=\ell/2$ in the sense that from that time onwards they lie completely in the \VBTZ part and will henceforth equal that of the eternal \VBTZ spacetime. The shell crossing geodesics generically cause the entropy to show a smooth interpolation between vacuum \ads and \VBTZ. We could call this a prethermalization, but note that the saturation value could be smaller than that of the vacuum due to the possibility of a locally negative energy density.\\

We have illustrated the behaviour of the entanglement entropy in two different cases. The first case is that of some generic oscillatory inhomogeneities modeled by some trigonometric functions, while the second case is that of a bilocal quench which is dual to a collapse of two point particles. The oscillatory quench shows a smooth interpolation from \ads to \VBTZ and an oscillatory behaviour in the final steady state, and the time it takes to reach the \VBTZ result is given by $\ell/2$ where $\ell$ is the length of the interval. The final state is reproduced quite well by perturbation theory to second order when the inhomogeneities are small. In the bilocal quench, the entanglement entropy is mainly governed by the location of the particle excitations compared to the location of the interval. Roughly speaking, in the \VBTZ spacetime, which is the late time geometry after the shell, the entropy is larger if there are more excitations inside the interval and is equal to that of \ads if there are no excitations inside the interval. During the prethermalization phase the entropy interpolates from that of \ads to that of \VBTZ during a time window which is at most $\ell/2$, but can be smaller since the energy density is zero on some parts of the shell. Another interesting feature is that during the prethermalization phase, the entanglement entropy is equal to that of \ads until at least one of the stress-energy tensor excitations has crossed one of the endpoints of the interval. Under specific circumstances there is also a phase transition in the sense that two geodesics exchange dominance and the evolution of the entropy is not smooth.\\

A second quantity of interest to us has been the differential entropy. It allows us to diagnoze the existence of an entanglement shadow and the possibility of bulk reconstruction via holographic entanglement entropy. We find that both in the homogeneous and in the inhomogeneous shell an entanglement shadow dynamically appears. Differential entropy smoothly interpolates between its value in \ads and that in eternal \VBTZ . In contrast with the entanglement entropy, the late time differential entropy is constant. The integration over the family of observers in some sense averages out the effect of the inhomogeneities.\\

In static geometries, it is known that long geodesics penetrate the entanglement shadow. Their length computes entwinement \cite{Balasubramanian:2014sra}, a field theoretic quantity that has been associated with the entanglement entropy of a subset of internal degrees of freedom in \cite{Balasubramanian:2016xho}. In homogeneous Vaidya long geodesics appear dynamically \cite{Ziogas:2015aja}. They probe the entanglement shadow, so it would be interesting to see if the notion of entwinement could also be applied to the infalling shell geometry. Moreover, in a string theory context such as the D1/D5 system thermalization has been associated with long string formation \cite{Carson:2014ena,Carson:2015ohj,Carson:2016uwf}. Long string formation arises in perturbation theory in the coupling of the D1/D5 CFT. Intuitively, it is tempting to associate the dynamical appearance of long geodesics with the entwinement of the long strings and it would be interesting to explore this analogy further.  In the inhomogeneous shell background, even the existence of long geodesics has not been studied yet.\\

We have restricted our study to inhomogeneous shells that are spinless and have a total mass above the black hole threshold. It would be interesting to extend this to shells with spin or to shells below the threshold. Below the threshold the physically relevant solution would be shells that bounce back at the center (similar to those in \cite{Bizon:2011gg}) instead of the more unphysical setup where the shell collapses to form a conical singularity. Studying entanglement entropies in such spacetimes would definitely be an interesting problem. Shells with angular momentum would be interesting since they would result in spacetimes where the stress-energy tensor modes $T_\pm$ are two different functions, but it is not known how to construct such spacetimes with a coordinate transformation analogous to \eqref{coord_tr} or how they can be constructed from a limit of infalling point particles.

\section*{Acknowledgements}
We would like to thank V. Balasubramanian, A. Bernamonti,  B. Craps and F. Galli for useful discussions on this subject and for early collaboration on the differential entropy in homogeneous Vaidya. We would also like to thank B. Craps for carefully reading the manuscript before publication. T.D.J would also like to thank NORDITA for hospitality while this work was in progress. This research has been supported in part
by FWO-Vlaanderen via projects G044016N and G006918N. T.D.J is Aspirant FWO-Vlaanderen.

\appendix

\section{Derivation of $F$ in the two-particle collision spacetime}\label{2p_appendix}

In this appendix we will derive the function $F$, which in our formulation completely specifies the spacetime, for the two-particle collision process. We will do this by starting with the standard formulation of this spacetime found in for instance  \cite{Matschull:1998rv}, and then performing a set of coordinate transformations. 

We will assume that the particles fall in from the boundary at angles $0$ and $\pi$, starting at time $t=-\pi/2$. These particles then collide at time $t=0$ at the center of \ads. The two particles' interaction with gravity is achieved by cutting out two moving regions connected to the particles that induce a holonomy around the worldlines of the particles. We will describe each of these regions as being bounded by two surfaces $w^\pm_1$ for particle 1 and $w^\pm_2$ for particle 2. The surface $w^+_i$ is identified with $w^-_i$ via an isometry of \ads. The surfaces $w^\pm_i$ are defined by
\begin{align}
w_1^\pm:& \tanh\tilde\chi\sin(-\tilde\vp\pm\epsilon)=\mp\sin\epsilon\sin \tilde t,\\
w_2^\pm:& \tanh\tilde\chi\sin(-\tilde\vp\pm\epsilon+\pi)=\mp\sin\epsilon\sin \tilde t.
\end{align}
The parameter $\epsilon$ can be interpreted as the energy of the particles. After the collision $(\tilde t>0)$, the resulting spacetime is now instead described by two regions, one region bounded by $w_{1,2}^-=w^+_1$ and $w^+_{1,2}=w^-_2$, and the other region bounded by $w_{2,1}^-=w^+_2$ and $w_{2,1}^+=w^-_1$, and this final spacetime can be interpreted as a black hole. The surfaces $w^\pm_{1,2}$ are described by the equations
\begin{equation}
\tanh\tilde\chi\sin(-\tilde\vp\mp\Gamma+\pi/2)=\pm\sin\Gamma\coth\xi\sin\tilde t,
\end{equation}
and the surfaces $w^\pm_{2,1}$ by
\begin{equation}
\tanh\tilde\chi\sin(-\tilde\vp\mp\Gamma-\pi/2)=\pm\sin\Gamma\coth\xi\sin\tilde t,\label{wedge2}
\end{equation}
The parameter $\Gamma$ is related to $\epsilon$ by $\Gamma=\epsilon-\pi/2$ and the parameter $\xi$ is determined by $\tanh\xi=-1/\tan\epsilon$. We also introduce a parameter $\mu$ by $\tan\Gamma=\tanh\mu\sinh\xi$ where $\mu$ will be related to the mass of the final black hole. The metric of \ads in these coordinates is given by $ds^2=-\cosh^2\tilde\chi d\tilde t^2+d\tilde\chi^2+\sinh^2\tilde\chi d\tilde\vp^2$. These coordinates are continuously connected to the AdS coordinates before the collision. We now want to map the late time geometry, as defined by these identifications, to the standard description of a black hole geometry given by equation \eqref{BTZ_metric} and use this to figure out what the transition function $F$ is when crossing the lightlike surface on which the particles are located. We will focus on the second wedge (centered around $-\pi/2$), the other wedge works similarly. We will first make a coordinate transformation to coordinates $t,\chi,\vp$ with the same metric, but such that the wedge takes the form \eqref{wedge2} but with $\xi=0$, namely
\begin{equation}
\tanh\chi\cos\vp=\mp\tanh\mu\sin t.
\end{equation}
The coordinate transformation to achieve this can be found in \cite{Lindgren:2015fum} and reads
\begin{align*}
\cos\tilde t\cosh\tilde\chi&=\cos t\cosh\chi,\\
\sin\tilde t\cosh\tilde\chi&=\cosh\chi\sin t\cosh\xi+\sinh\chi\sinh\xi\sin\vp,\\
\sinh\tilde\chi\cos\tilde\vp&=\sinh\chi\cos\vp,\\
\sinh\tilde\chi\sin\tilde\vp&= \cosh\chi\sinh\xi\sin t+\sinh\chi\cosh\xi\sin\vp.
\end{align*}
The metric in the non-tilde coordinates is the same \ads metric as the tilde coordinates (in other words, this coordinate transformation is an isometry). Now we want to bring this to coordinates $\sigma,\rho,y$ which are such that the metric takes the form $-\frac{1}{-1+\rho^2}d\sigma^2+(-1+\rho^2)d\rho^2+\rho^2dy^2$. In these coordinates the surfaces $w_{2,1}^\pm$ will be mapped to surfaces at constant $y$. To find this coordinate transformation, note that for the AdS coordinate system we can use the embedding equation for \ads, $x_3^2+x_0^2-x_1^2-x_2^2=1$. The coordinates $t,\chi,\vp$ can be obtained from
\begin{equation}
x^0=\cosh\chi\sin t,\quad x^1=\sinh\chi\cos\vp,\quad x^2=\sinh\chi\sin\vp,\quad x^3=\cosh\chi\cos t,
\end{equation}
while the black hole coordinates $\sigma,\rho,y$ can be obtained from
\begin{equation}
x^0=-\cosh\beta\cosh y,\quad x^1=\cosh\beta\sinh y,\quad x^2=\sinh\beta\cosh \sigma,\quad x^3=\sinh\beta\sinh\sigma,
\end{equation}
where $\rho=\cosh\beta$. By comparing these two embeddings we obtain our coordinate transformation. It is easy to see now that the two surfaces $w_{2,1}^\pm$ corrsponds to $y=\pm \mu$. We are now interested in figuring out how the lightlike surface, on which the particles fall in on, is transformed under this coordinate transformation. A lightray on this surface can be defined by an angle $\vp=\psi$ and the relation $\tanh\chi=-\sin t$. Formulating it in terms of the $x^i$ coordinates, we see that we have $x_1^2+x_2^2=x_0^2$ as well as $x^1=-x^0\cos\psi$. Thus we have 
\begin{equation}
\rho^2=x_0^2-x_1^2=x_2^2=\sinh^2\chi\sin^2\psi,\quad \tanh y=\cos\psi.
\end{equation}
Furthermore, between the tilde coordinates and the non-tilde AdS coordinates, we obtain the relations 
\begin{align}
\sinh\chi\cos\psi=&\sinh\tilde\chi\cos\tilde\psi,\quad \cos\psi=\frac{\cos\tilde\psi}{\cosh\xi+\sin\tilde\psi\sinh\xi},\nonumber\\ \sin\psi=&\frac{\sinh\xi+\cosh\xi\sin\tilde\psi}{\cosh\xi+\sinh\xi\sin\tilde\psi}.
\end{align}
From this we can finally obtain
\begin{equation}
\rho=-\sinh\tilde\chi(\cosh\xi\sin\tilde\psi+\sinh\xi),\quad \tanh y=\frac{\cos\tilde\psi}{\cosh\xi+\sinh\xi\sin\tilde\psi}
\end{equation}
Note that $\xi<0$. Now it can be shown that if we write $y=H(\tilde\vp)$ we indeed have $\rho^2=r^2/(H'(\tilde\vp))^2$ which is required by consistency in our formalism. The borders of the wedge are now located at $y=\pm\mu$, and thus by considering $\tilde\psi=0,-\pi$ we also have the relation $\tanh\mu=1/\cosh\xi$. Note that we only worked with one wedge so far, and thus the total range of the coordinate $y$ is $4\mu$. Thus in the correct rescaled coordinates where the metric takes the form $-\frac{1}{-M+\rho^2}d\sigma^2+(-M+\rho^2)d\rho^2+\rho^2dy^2$, with $M=(4\mu)^2/4\pi^2$ we have $H'(\tilde\vp)=-\frac{1}{\sqrt{M}} \frac{1}{\cosh\xi\sin\tilde\vp+\sinh\xi}$. Gluing together the two wedges, the full transition function $F$ is thus given by $F(\tilde\vp)=H(\tilde\vp)$ for $-\pi<\tilde\vp<0$ (which maps to the interval $(-\pi/2,\pi/2)$) and $F(\tilde\vp)=H(\tilde\vp-\pi)+\pi$ for $0<\tilde\vp<\pi$ (which maps to the interval $(\pi/2,3\pi/2)$). Note that the stress-energy tensor is constant and equal to $-1/4$ everywhere except at $\tilde\vp=0,\pi$ where it has a deltafunction with strength $-\coth\xi=\tan\epsilon$ which comes from the third derivative of $F$ in the stress-energy tensor, which is in agreement with the notion that $\epsilon$ is a measure of the energy of the point particles. Note that in section \ref{bilocal} we rotate the setup such that the particles fall in at $\pm\pi/2$ instead. In this case the (derivative of) the transition function, after also expression $\xi$ and $\mu$ in terms of $M$, is given by
\begin{equation}
F'(\tilde\vp)= \Bigg\{
\begin{array}{cc}
\frac{1}{\sqrt{M}} \frac{\sinh\LF \frac{\pi \sqrt{M}}{2}\RF}{\cosh\LF \frac{\pi \sqrt{M}}{2}\RF\cos\LF \tilde\vp \RF +1},& -\pi/2\leq\tilde\vp\leq\pi/2 \\
                       \frac{1}{\sqrt{M}} \frac{\sinh\LF \frac{\pi \sqrt{M}}{2}\RF}{-\cosh\LF \frac{\pi \sqrt{M}}{2}\RF\cos\LF \tilde\vp \RF +1},&\pi/2\leq\tilde\vp\leq3\pi/2\\
\end{array}
\end{equation}
and extended periodically to all other angles.

\section{Inhomogeneous refraction conditions}\label{app:inhom_refrac}
The refraction conditions on the inhomogeneous shell can be derived from a variation of the geodesic action
\begin{equation}
\mathcal{I}= \int\limits_{\la_1}^{\la_{s_1}} \sqrt{g_{\mu\nu}^{>} \dot{x}^\mu \dot{x}^\nu} d\la +  \int\limits_{\la_{s_1}}^{\la_{s_2}} \sqrt{g_{\mu\nu}^{<} \dot{x}^\mu \dot{x}^\nu} d\la +  \int\limits_{\la_{s_2}}^{\la_2} \sqrt{g_{\mu\nu}^{>} \dot{x}^\mu \dot{x}^\nu} d\la,
\end{equation}
where we have explicitly split the geodesic into its two legs in the \VBTZ  geometry at $v>0$ and a leg in the \ads geometry at $v<0$. Upon using the geodesic equations of motion in each patch, the variation of the action becomes a sum of boundary terms
\begin{equation}
\delta \mathcal{I} = \LT g_{\mu\nu}^{>} \dot{x}^\mu \delta x^\nu \RT^{\la_{s_1}}_{\la_1} + \LT g_{\mu\nu}^{<} \dot{x}^\mu \delta x^\nu\RT^{\la_{s_2}}_{\la_{s_1}}
+ \LT g_{\mu\nu}^{>} \dot{x}^\mu \delta x^\nu\RT^{\la_2}_{\la_{s_2}}.
\end{equation}
The variation on the asymptotic boundary is fixed so $\delta x^\nu \LF \la_i\RF=0$. Now we pick $(r,\phim,v)$ coordinates on \ads and $(\rp,\phip,v)$ coordinates on \VBTZ which is mapped to static BTZ by this choice of coordinates. First of all note that the shell is located at $v=0$, so $\delta v(\la_{s_i})=0$. The first order action becomes
\begin{align}
\delta \mathcal{I} = &\dot{v}_{\text{BTZ}}(\la_{s_1})\delta \rp(\la_{s_1}) - \dot{v}_{\text{\ads}}(\la_{s_1})\delta r(\la_{s_1}) + 2\rp^2\dot{\phip}(\la_{s_1}) \delta\phip(\la_{s_1})\nonumber\\
& -2r^2\dot{\vp}(\la_{s_1})\delta \vp(\la_{s_1}) + \LF \la_{s_1}\leftrightarrow \la_{s_2}\RF.
\end{align}
The variations of $\delta \bar{r}$ and $\delta \phip$ are linked to that of $\delta r$, $\delta\phim$ by the coordinate transformation on the shell~(\ref{coord_tr}). Then we use that the variation at $\la=\la_{s_1}$ is independent from that at $\la=\la_{s_2}$ and that variations in $r$ are independent from variations of $\vp$. Given that $\phip=F(\vp)$ on the shell, we will define the quantity $\dot{\vp}_{\text{BTZ}}=\dot{\phip}/F'(\vp)$ evaluated on the shell. Demanding that the variation of the action should vanish implies the conditions
\begin{align}
\frac{\delta \mathcal{I}}{\delta \vp\LF \la_{s_i}\RF} &= 2\dot{\vp}_{\text{BTZ}} r^2 - 2r\dot{v}_{\text{BTZ}} \frac{F^{''}\LF \vp\RF}{F^{'2}\LF \vp\RF} - 2r^2\dot{\vp}_{\text{\ads}}=0,\\
\frac{\delta \mathcal{I}}{\delta r\LF \la_{s_i}\RF} &= \frac{\dot{v}_{\text{BTZ}}}{F^{'}\LF \vp\RF} - \dot{v}_{\text{\ads}}=0,
\end{align}
both for $i=1,2$ and where all quantities are evaluated at the shell. When the shell is homogeneous, then $F'\LF \vp\RF=1$ and the refraction conditions reduce to \eqref{eq:refrac_hom}. In this derivation we have assumed that the affine parameter $\lambda$ is continuous across the shell, and in particular it runs monotonically along the geodesic. If the affine parameter instead would change direction when crossing the shell one would need to change the signs of either the \ads or the BTZ terms in the refraction conditions above.

\section{Saturation time}\label{app:saturation}
In this appendix we will prove that the geodesic will stop intersecting the shell at exactly $t=\ell/2$. This will typically be the saturation time (the time after which the entanglement entropy will be equal to that of the \VBTZ geometry), but the entropy can also saturate earlier if the shell has zero energy density in some regions as is the case in the bilocal quench in section \ref{bilocal}.\\

To find the time where the geodesic no longer intersects the shell, we start with the geodesic in the late time \VBTZ part of the spacetime. We then show that at $t=\ell/2$, the geodesic will be exactly tangent to the shell at $v=0$. From equation \eqref{eq:EOM_r} and \eqref{eq:EOM_v} we obtain that $\dot{v}=0$ implies that $r=|J| r_H$. The other possibility, that $r=r_H$, is excluded since the geodesic lies outside the horizon. It is possible to express $v$ in terms of $r$ to obtain\cite{Balasubramanian:2011ur}
\begin{equation}
v(r)=v_0+\frac{1}{2r_H}\log\LF\frac{r-r_H}{r+r_H}\frac{r^2-(E+1)r_H^2+r\dot{r}}{r^2+(E-1)r_H^2+r\dot{r}}\RF.\label{vr}
\end{equation}
Note that for $r=|J| r_H$, we have $\dot{r}=\pm |E|r_H$ where the sign corresponds to the two solutions of $\lambda$. However, note that from \eqref{eq:EOM_v}, for $\dot{v}$ to vanish, we must require that $\dot{r}$ has the opposite sign to $E$ and thus we pick the solution for $\lambda$ that satisfies $\dot{r}=-E r_H$. Using \eqref{eq:E_pm} and \eqref{eq:J_pm} for $E$ and $J$ we obtain that the minimum value of $v$ is given by
\begin{equation}
v_{\text{min}}=v_0+\frac{1}{2}(a+b)=\frac{1}{2}\LF F(t+\zeta)-F(-t+\zeta+\ell)\RF,\label{vmin_eq}
\end{equation}
where we have then used \eqref{eq:EJeq1} and \eqref{eq:EJeq2} to express $a$ and $b$ in terms of the function $F$, the boundary time $t$, the location of the interval $\zeta$ and the interval size $\ell$. Now we see that $v_{\text{min}}=0$ for $t=\ell/2$. Moreover, since $F$ is a monotonically increasing function, we see that the right hand side of \eqref{vmin_eq} is increasing as well and thus we conclude that the geodesic does not intersect the shell for $t>\ell/2$ and must intersect the shell for $t<\ell/2$.


\providecommand{\href}[2]{#2}\begingroup\raggedright\endgroup

\end{document}